\documentclass[useAMS,usenatbib]{mn2e}

\input psfig.sty
\usepackage{xspace}
\usepackage{graphicx}
\usepackage{amsmath,amssymb}

\newcommand {\apj} {ApJ}
\newcommand {\apjl} {ApJL}
\newcommand {\apjs} {ApJS}
\newcommand {\mnras} {MNRAS}

\newcommand {\aap} {A\&A}
\newcommand {\aj} {AJ}

\newcommand {\nat} {Nature}

\newcommand {\pasp} {PASP}

\newcommand {\etal} {et~al.~}
\def \spose#1{\hbox  to 0pt{#1\hss}}  
\newcommand {\lta} {\mathrel{\spose{\lower 3pt\hbox{$\sim$}}\raise  2.0pt\hbox{$<$}}}
\newcommand {\gta} {\mathrel{\spose{\lower  3pt\hbox{$\sim$}}\raise 2.0pt\hbox{$>$}}}

\newcommand {\ha}  {\ifmmode H\alpha \else H$\alpha $ \fi} 

\def\Sref#1{Section~\ref{#1}\xspace}
\def\Fref#1{Figure~\ref{#1}\xspace}
\def\Tref#1{Table~\ref{#1}\xspace}
\def\Eref#1{Equation~\ref{#1}\xspace}

\newcommand {\kms} {\ifmmode  \,\rm km\,s^{-1} \else $\,\rm km\,s^{-1}  $ \fi }
\newcommand {\kpc} {\ifmmode  {\rm kpc}  \else ${\rm  kpc}$ \fi  }  
\newcommand {\pc} {\ifmmode  {\rm pc}  \else ${\rm pc}$ \fi  }  
\newcommand {\Msun} {\ifmmode {\rm M_{\odot}} \else ${\rm M_{\odot}}$ \fi} 
\newcommand {\Zsun} {\ifmmode {\rm Z_{\odot}} \else ${\rm Z_{\odot}}$ \fi} 
\newcommand {\yr} {\ifmmode yr^{-1} \else $yr^{-1}$ \fi} 
\newcommand {\hMsun} {\ifmmode h^{-1}\,\rm M_{\odot} \else $h^{-1}\,\rm M_{\odot}$ \fi}

\def\zd{z_{\rm d}}
\def\zs{z_{\rm s}}

\def\Dd{D_{\rm d}}
\def\Ds{D_{\rm s}}
\def\Dds{D_{\rm ds}}
\def\Sigmacrit{\Sigma_{\rm crit}}

\def\Vhalo{V_{\rm c,h}}
\def\rhalo{r_{\rm c,h}}
\def\qhalo{q_{3,\rm h}}
\def\vc{V_{\rm c}}
\def\rc{r_{\rm c}}
\def\q3{q_{3}}
\def\bsis{b_{\rm SIS}}

\def\Mstar{M_{*}}
\def\logMstar{\log_{10}\left(\Mstar/\Msun\right)}
\def\Mstarb{M_{*,\rm b}}

\def\Mstard{M_{*,\rm d}}

\def\Reff{R_{\rm 50}}
\def\Reffb{R_{\rm 50,b}}
\def\Reffd{R_{\rm 50,d}}
\def\sersic{S\'ersic}
\def\nb{n_{\rm b}}
\def\qb{q_{\rm b}}

\def\qd{q_{\rm d}}

\def\hst{{\it HST}\xspace}

\def\python{{\sc python}\xspace}

\def\Bfilter{F450W\xspace}
\def\Vfilter{F606W\xspace}
\def\Ifilter{F814W\xspace}
\def\Kfilter{K'\xspace}

\def\Kband{\Kfilter-band\xspace}

\def\pr{{\rm Pr}}
\def\data{{\mathbf{d}}}
\def\datap{{\mathbf{d}^{\rm p}}}
\def\datai{d_i}
\def\datapi{d^{\rm p}_i}
\def\masspars{\boldsymbol{\theta}_{\rm m}}
\def\srcpars{\boldsymbol{\theta}_{\rm s}}
\def\vrot{{\mathbf{v}}}
\def\vrotp{{\mathbf{v}^{\rm p}}}
\def\vrotmodel{\hat{\mathbf{v}}}
\def\vrotj{v_j}
\def\vrotpj{v^{\rm p}_j}

\def\Angstrom{A\xspace}

\def\NaD{Na\,{\sc D}\xspace} 
\def\NII{N\,{\sc ii}\xspace}

\def\Mgb{Mg\,{\rm b}\xspace}
\def\Ha{H$\alpha$\xspace}

\def\OII{[O\,{\sc ii}]\xspace}

\def\FeII{[Fe\,{\sc ii}]\xspace}

\def\NSWELLS{20}

\def\uvic{Dept. of Physics and Astronomy, 
  University of Victoria, Victoria, BC, V8P 5C2, Canada}
\def\lick{UCO/Lick Observatory, 
  University of California, Santa Cruz, CA 95064, USA}
\def\ucsb{Dept. of Physics, University of California, 
  Santa Barbara, CA 93106, USA}
\def\kipac{Kavli Institue for Particle Astrophysics and Cosmology, 
  P.O. Box 20450, MS29, Stanford, CA 94309, USA}
\def\utah{Department of Physics and Astronomy, University of Utah, 
  Salt Lake City, UT 84112, USA}
\def\kapteyn{Kapteyn Astronomical Institute, University of Groningen, 
  P.O.Box 800, 9700 AV Groningen, The Netherlands}

\def\duttonemail{\tt dutton@uvic.ca}

\def\packard{Packard Research Fellow}
\def\cita{CITA National Fellow}

\newcommand{\lens}{{SDSS\,J2141$-$0001}\xspace}

\title [The disk and halo of lens \lens] {The SWELLS survey. II.\\
  Breaking the disk-halo degeneracy in the spiral galaxy gravitational
  lens \lens\thanks{Based in part on observations made with the NASA
  / ESA Hubble Space Telescope, obtained at the Space Telescope
  Science Institute, which is operated by AURA, Inc., under NASA
  contract NAS 5-26555. These observations are associated with
  programs 10587 and 11978.}.}

\author[Dutton \etal]{%
  Aaron~A.~Dutton$^{1,2}$\thanks{\duttonemail}\thanks{\cita},
  Brendon~J.~Brewer$^{3}$, 
  Philip~J.~Marshall$^{3,4}$,
  M.~W.~Auger$^3$,
\newauthor{%
  Tommaso~Treu$^3$\thanks{\packard}, 
  David~C.~Koo$^2$, 
  Adam~S.~Bolton$^5$, 
  Bradford~P.~Holden$^2$,}
\newauthor{%
  Leon~V.~E.~Koopmans$^6$}\\
  $^1$\uvic\\
  $^2$\lick\\
  $^3$\ucsb\\
  $^4$\kipac\\
  $^5$\utah\\
  $^6$\kapteyn
}

\begin{document}
             
\date{accepted to MNRAS}
             
\pagerange{\pageref{firstpage}--\pageref{lastpage}}\pubyear{2011}

\maketitle         

\label{firstpage}


\begin{abstract}
  The degeneracy among the disk, bulge and halo contributions to
  galaxy rotation curves prevents an understanding of the distribution
  of baryons and dark matter in disk galaxies.  In an attempt to break
  this degeneracy, we present an analysis of the strong gravitational
  lens \lens, discovered as part of the SLACS survey. The lens galaxy
  is a high inclination, disk dominated system.  We present new Hubble
  Space Telescope multicolor imaging, gas and stellar kinematics data
  derived from long-slit spectroscopy, and K-band laser guide star
  adaptive optics imaging, both from the Keck telescopes. We model the
  galaxy as a sum of concentric axisymmetric bulge, disk and halo
  components and infer the contribution of each component, using
  information from gravitational lensing and gas kinematics.  This
  analysis yields a best-fitting total (disk plus bulge) stellar mass
  of $\log_{10}(\Mstar/\Msun)=10.99^{+0.11}_{-0.25}$. The photometric
  data combined with stellar population synthesis models yield
  $\log_{10}(\Mstar/\Msun)=10.97\pm 0.07$, and $11.21\pm 0.07$ for the
  Chabrier and Salpeter IMFs, respectively. Assuming no cold gas, a
  Salpeter IMF is marginally disfavored, with a Bayes factor of
  2.7. Accounting for the expected gas fraction of $\simeq 20\%$
  reduces the lensing plus kinematics stellar mass by $0.10\pm0.05$
  dex, resulting in a Bayes factor of 11.9 in favor of a Chabrier IMF.
  The dark matter halo is roughly spherical, with minor to major axis
  ratio $\qhalo=0.91^{+0.15}_{-0.13}$.  The dark matter halo has a
  maximum circular velocity of $V_{\rm max}=276^{+17}_{-18} \kms$, and
  a central density parameter of
  $\log_{10}\Delta_{V/2}=5.9^{+0.9}_{-0.5}$. This is higher than
  predicted for uncontracted dark matter haloes in $\Lambda$CDM
  cosmologies, $\log_{10}\Delta_{V/2}=5.2$, suggesting that either the
  halo has contracted in response to galaxy formation, or that the
  halo has a higher than average concentration. Larger samples of
  spiral galaxy strong gravitational lenses are needed in order to
  distinguish between these two possibilities.  At 2.2 disk scale
  lengths the dark matter fraction is $f_{\rm DM}=0.55^{+0.20}_{-0.15}$, 
  suggesting that \lens is sub-maximal.
\end{abstract}


\setcounter{footnote}{1}


\section{Introduction}
\label{sec:intro}

\begin{figure*}
  \centering\includegraphics[width=0.9\linewidth]{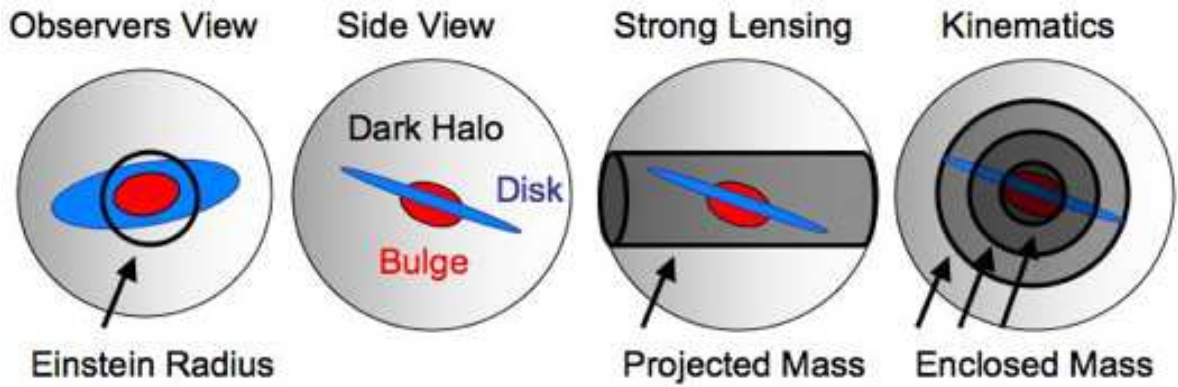}
\caption{Illustration of the different geometries probed by strong
  lensing and kinematics. Strong lensing measures mass with a cylinder
  (or more generally an ellipse), whereas stellar and gas kinematics
  measure mass within spheres (or more generally ellipsoids).}
\label{fig:cartoon}
\end{figure*}

The discovery of extended flat rotation curves in the outer parts of
disk galaxies three decades ago (Bosma 1978; Rubin \etal 1978) was
decisive in ushering in the paradigm shift that led to the now
standard cosmological model dominated by cold dark matter (CDM).  The
need for dark matter on cosmological scales is also firmly established
from observations of the Cosmic Microwave Background, type Ia
Supernovae, weak lensing, and galaxy clustering (see, e.g., Spergel
\etal 2007).  Numerical simulations of structure formation within the
$\Lambda$CDM cosmology make firm predictions for the structure and
mass function of dark matter haloes in the absence of baryons (e.g.,
Navarro, Frenk, \& White 1997; Bullock \etal 2001; Macci\`o \etal
2007; Navarro et al.\ 2010).

It is still unclear, however, whether this standard model can
reproduce the observed properties of the Universe at galactic and
sub-galactic scales.  There are problems related to the inner density
profiles of dark matter haloes (e.g., de Blok \etal 2001; Swaters
\etal 2003; Newman et al. 2009), reproducing the zero point of the
Tully-Fisher relation (e.g., Mo \& Mao 2000; Dutton \etal 2007), and
the amount of small-scale substructure (e.g., Klypin \etal 1999; Moore
et al. 1999; Stewart \etal 2008).  There are three classes of
solutions to these problems: those that invoke galaxy formation
processes that modify the properties of dark matter haloes; those that
change the nature of dark matter itself; and those in which dark
matter does not exist.  Thus, measuring the density profiles of the
dark matter haloes of galaxies of all types is a stringent test for
galaxy formation theories.

From an observational point of view, little is known about the
detailed distribution of dark matter in the inner regions of disk
galaxies, despite the great investment of telescope time and high
quality measurements of hundreds of rotation curves (e.g., Carignan \&
Freeman 1985; Begeman 1987; Courteau 1997; de Blok \& McGaugh 1997;
Verheijen 1997; Swaters 1999; de Blok \etal 2001; Swaters \etal 2003;
Blais-Ouellette \etal 2004; Simon \etal 2005; Noordermeer \etal 2005;
Simon \etal 2005; Chemin \etal 2006; Kuzio de Naray 2006; de Blok
\etal 2008; Dicaire \etal 2008; Epinat \etal 2008). The fundamental
reason is the so-called {\it disk-halo degeneracy}: mass models with
either maximal or minimal baryonic components fit the rotation curves
equally well, leaving the structure of the dark matter halo poorly
constrained by the kinematic data alone (e.g., van Albada \& Sancisi
1986; van den Bosch \& Swaters 2001; Dutton \etal 2005).  Stellar
population models are able to place constraints on stellar
mass-to-light ratios, allowing inference about the baryonic
contribution to the overall mass profile. However, there are a number
of uncertainties which limit the accuracy of this method (e.g., Conroy
\etal 2009, 2010). These include systematic uncertainties such as the
unknown stellar initial mass function (IMF), and the treatment of the
various stellar evolutionary phases in stellar population synthesis
(SPS) models. These result in about a factor of 2 uncertainty in the
stellar masses estimated from spectral energy distribution (SED)
fitting. Moreover, for a given IMF and SPS model, there are
uncertainties in the star formation histories, metallicities and
extinction which introduce ($1\sigma$) random errors in measurements
of stellar masses for individual galaxies at the level of 0.15~dex
(e.g., Bell \& de Jong 2001; Auger \etal 2009, Gallazzi \& Bell 2009).

Nevertheless, galaxy colours and dynamical mass estimates have been
used in combination by various authors to place an upper limit on the
stellar mass-to-light ratio normalisation, favoring IMFs more bottom
light than Salpeter for spiral galaxies and fast rotating low-mass
elliptical galaxies (Bell \& de Jong 2001; Cappellari \etal 2006; de
Jong \& Bell 2007; see, however, Treu et al. 2010, Auger et al. 2010,
and van Dokkum \& Conroy 2010 for massive ellipticals).  However, as
theory generally predicts more dark matter in the inner regions of
disk galaxies than is consistent with standard IMFs (Dutton \etal
2007; Dutton \etal 2010b) a lower limit to the stellar mass would
provide a more useful constraint for $\Lambda$CDM.

Several other methods have been used to try and measure disk galaxy
stellar masses, independent of the uncertainties in the IMF.  These
include: 1) vertical velocity dispersions of low inclination disks
(Bottema 1993; Verheijen \etal 2007; Bershady \etal 2010), 2) bars and
spiral structure (Weiner \etal 2001; Kranz \etal 2003), and 3) strong
gravitational lensing by inclined disks (Maller \etal 2000; Winn \etal
2003).  None of these methods have thus far yielded conclusive
results.

\begin{figure*}
  \centering\includegraphics[width=0.42\linewidth]
  {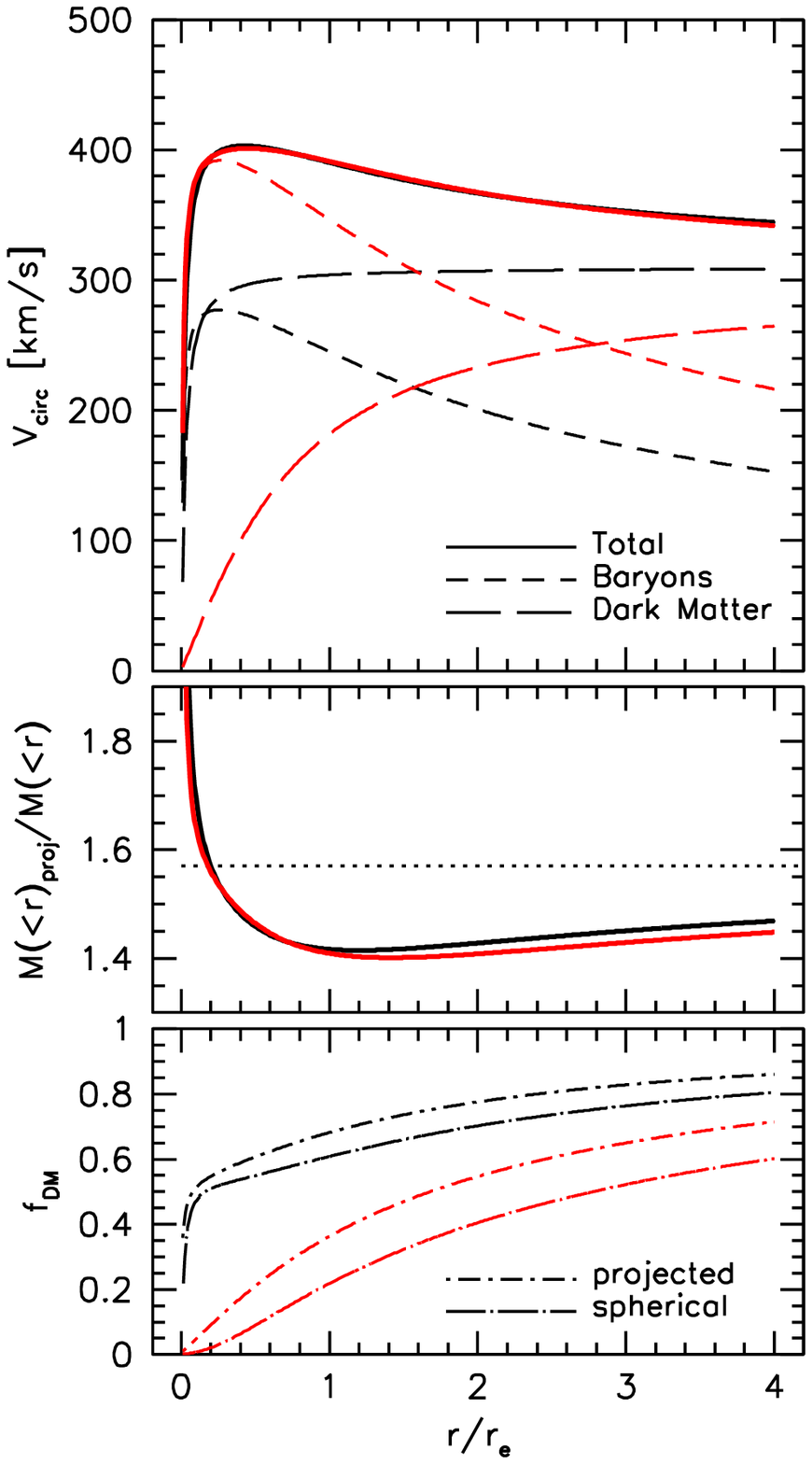}
  \centering\includegraphics[width=0.42\linewidth]
  {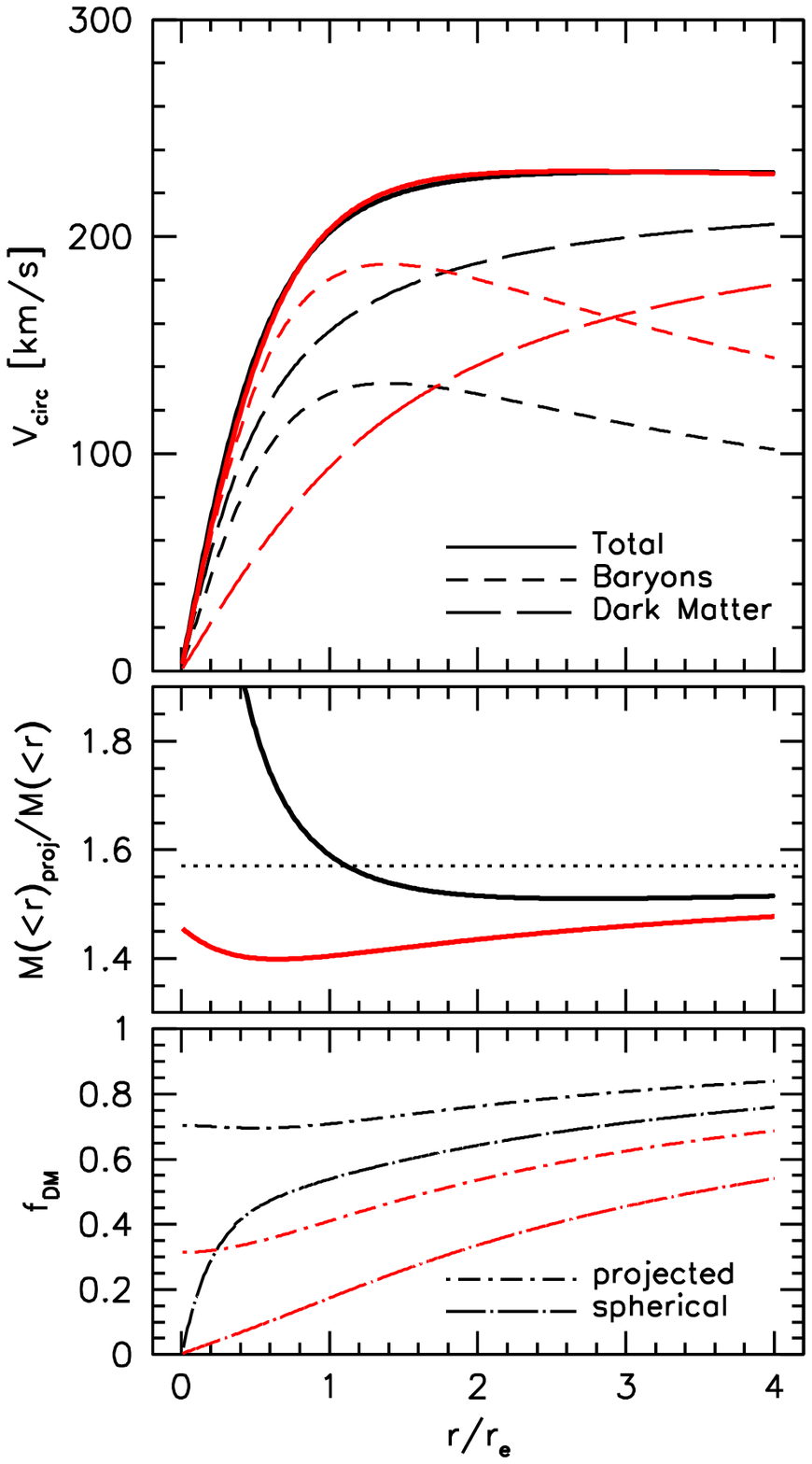}
  \caption{Differences between projected (cylindrical) mass and
    enclosed (spherical) mass for a bulge-halo system (left) and a
    disk-halo system (right). For each system two models are
      shown (in red and black). The models have baryonic mass profiles
      (short-dashed lines, upper panels) with the same shape but
      normalizations that differ by a factor of two. The dark matter
      profiles (long-dashed lines, upper panels) have been chosen so
      that the total circular velocity curves are close to identical
      (solid lines, upper panels). For the bulge-halo system the
    ratio between projected and enclosed masses (middle panels) is
    independent of the relative contributions of the bulge and halo,
    which differ significantly between the two models (lower
    panels). However, for the disk-halo system there is a significant
    difference between the projected and enclosed masses, especially
    at radii smaller than the effective radius. This illustrates the
    potential of strong lensing plus kinematics to break the disk-halo
    degeneracy.}
\label{fig:proj}
\end{figure*}

An approach combining strong gravitational lensing plus kinematics
holds great promise, because it takes advantage of the different
geometries of disks and haloes, which results in three effects that
enable the disk mass to be measured. 1) An inclined disk will present
a much higher projected surface density than a face-on disk, with
resulting image positions and shapes that depend on the disk mass
fraction.  2) An edge-on disk is highly elliptical in projection, more
than expected for any realistic dark matter halo, with resulting total
mass ellipticity depending on the disk mass fraction 3) Strong lensing
measures mass projected along a cylinder (within the Einstein radius),
whereas stellar kinematics (rotation and dispersion) measure mass
enclosed within spheres (see \Fref{fig:cartoon}). For spherical mass
distributions of stars and dark matter, the ratio between the
projected mass within a cylinder of radius, $r$, and the enclosed mass
within a sphere of the same radius, $r$, is {\it independent} of the
relative contribution of the two mass components (left panel
\Fref{fig:proj}). Therefore, in order to break the degeneracy one has
to assume a radial profile shape for both components (e.g., Treu \&
Koopmans 2002, 2004; Koopmans \& Treu 2003; Koopmans et al. 2006; Treu
et al.\ 2010; Auger et al. 2010). Typically this involves assuming the
baryonic mass follows the light, and then assuming a functional form
for the dark matter halo. However, for a disk plus halo system, this
ratio is {\it dependent} on the relative contribution of the two
components (right panel \Fref{fig:proj}).  Thus if the spherical and
cylindrical masses can be measured accurately enough, the disk halo
degeneracy can be broken without assuming a specific radial profile
shape for either component. Furthermore, strong lensing plus
kinematics can place constraints on the 3D shape of the dark matter
halo (e.g., Koopmans, de Bruyn, \& Jackson 1998; Maller \etal 2000)
which is of interest because $\Lambda$CDM haloes are predicted to be
non-spherical (e.g. Allgood \etal 2006; Bett \etal 2007; Macci\`o
\etal 2008).

The power of the strong gravitational lensing method has not yet been
fully realised, primarily due to the scarcity of known spiral galaxy
gravitational lenses.  Prior to the SLACS Survey (Bolton \etal 2006,
2008) only a handful of spiral galaxy lenses with suitable
inclinations to enable rotation curve measurements were known:
Q2237$+$0305 (Huchra \etal 1985; Trott \& Webster 2002); B1600$+$434
(Jackson \etal 1995; Jaunsen \& Hjorth 1997); PMN\,J2004$-$1349 (Winn
\etal 2003); CXOCY\,J220132.8$-$320144 (Castander \etal 2006).
However, most of these systems are doubly-imaged QSOs which provide
minimal constraints on the projected mass density. Q2237$+$0305 is a
quadruply-imaged QSO, which gives more robust constraints, but since
the Einstein radius is small compared to the size of the galaxy, the
lensing is mostly sensitive to the bulge mass, not the halo (Trott et
al. 2010; van de Ven et al. 2010).

The final SLACS lens sample (Auger \etal 2009) is comprised of 98
strong galaxy-galaxy lenses, among these, 16 have been classified
morphologically as type S or S0.  Inspired by this, we have extended
the Sloan Digital Sky Survey (hereafter SDSS, York \etal 2000)
spectroscopic lens selection technique specifically to spiral galaxy
lenses. In the resulting SWELLS survey (Treu \etal, in prep, referred
to hereafter as Paper I) we have assembled a larger sample of \NSWELLS
\, late-type galaxy-scale gravitational lenses for detailed mass
modelling.  In this paper, the second of the SWELLS series, we present
a detailed and self-consistent mass model of the spiral galaxy lens
\lens (RA=21:41:54.67, DEC=$-$00:01:12.2, J2000), constrained by both
kinematic and lensing data. As we will see, this galaxy is
disk-dominated, with a disk inclination of $\simeq 80^\circ$; this
set-up approximately maximises the projected disk mass while allowing
an accurate rotation curve to be measured.

The original spectroscopic observations of \lens were obtained on SDSS
plate~989, with fiber~35, on MJD~52468. The latest public SDSS-DR7
(Abazajian \etal 2009) Petrosian magnitudes (uncorrected for
extinction) for the lens galaxy are
$(u,g,r,i,z)=(20.61,18.62,17.47,16.92,16.48)$ with errors
$(0.15,0.01,0.01,0.01,0.02)$. The SDSS measured redshift for the lens
galaxy is $\zd=0.1380 \pm 0.00015$, and the velocity dispersion is
$181\pm14\kms$.  The spectrum also exhibits nebular emission lines at
a background redshift of $\zs=0.7127$ (Bolton \etal 2008).  With these
redshifts the scale in the lens plane is $1\,{\rm arcsec}=2.438 \rm
\,kpc$, while in the source plane it is $1\,{\rm arcsec}=7.196 \rm
\,kpc$.

This paper is organised as follows.  In \Sref{sec:imaging} we present
the imaging observations of \lens from Keck and the Hubble Space
Telescope, and then infer the structure of the stellar mass
distribution of the galaxy in the presence of its dust from these data
in \Sref{sec:mstar}. With this information in hand we then define a
three-component mass model for the galaxy in \Sref{sec:model}, and
describe how it is constrained by the imaging data (although we choose
not to use the stellar mass inferred from the SED at this stage).  In
\Sref{sec:lens}, we describe the preparation and analysis of the
strong lensing data.  In \Sref{sec:spec} we present the spectroscopic
observations of \lens from Keck. Then in \Sref{sec:results} we present
fits to the lensing and kinematics data using three combinations:
lensing only; kinematics only; and lensing plus kinematics.  This
joint analysis yields constraints on the stellar mass of the disk and
bulge; returning to the stellar masses inferred from the stellar
population modelling of the SED we, discuss implications of our
results for the stellar IMF in \Sref{sec:imf}. In
\Sref{sec:halodensity} \& \Sref{sec:haloshape} we discuss our results
for the density and shape of the dark matter halo.  We conclude in
\Sref{sec:concl}.

Throughout, we assume a flat $\Lambda$CDM cosmology with present day
matter density, $\Omega_{\rm m}=0.3$, and Hubble parameter, $H_0=70
\rm\,km\,s^{-1}\,Mpc$.  All magnitudes are given in the AB system.
Unless otherwise stated, all parameter estimates are the median of the
marginalised posterior PDF, and their uncertainties are described by
the absolute difference between the median and the 84th and 16th
percentiles (such that the error bars enclose 68\% of the posterior
probability).


\begin{figure*}
\centering\includegraphics[width=0.95\linewidth]
{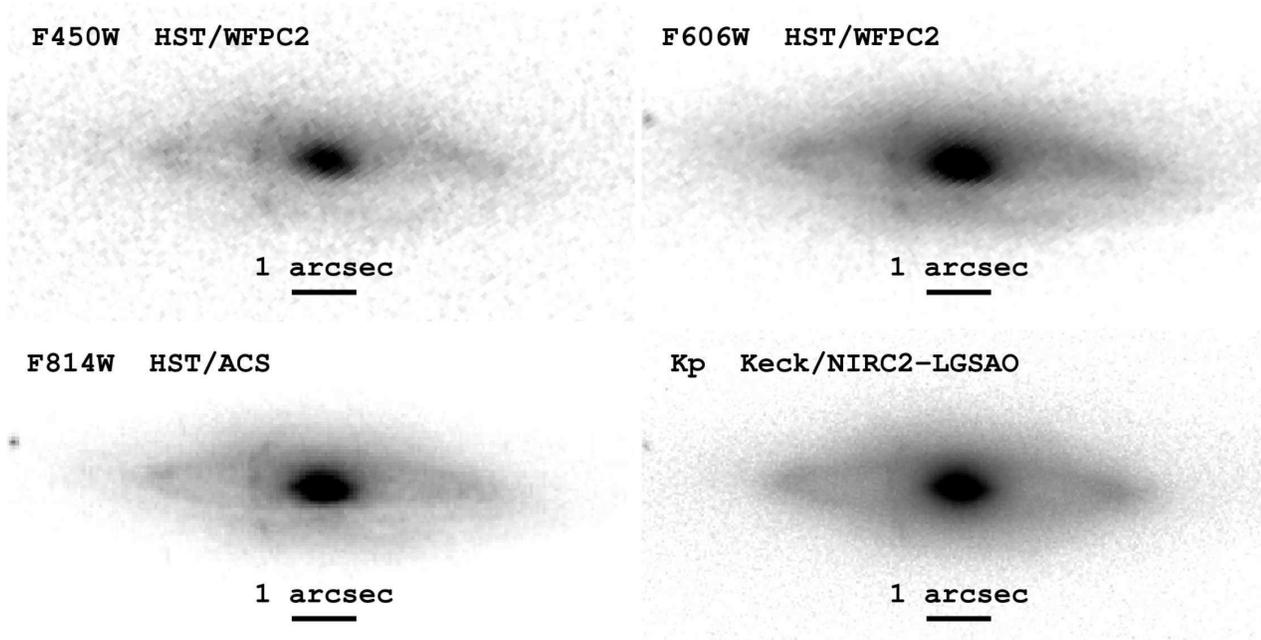}
\caption{Optical to near-IR high resolution imaging of \lens. Images
  are 12~arcsec by 6~arcsec, North up, East is left.  The lens is a
  high inclination, disk dominated, and star forming spiral
  galaxy. The source ($\simeq 1.3$ arcsec to the east of the galaxy
  center) appears to be multiply imaged in the optical, but is a
  continuous arc in the near IR.}
\label{fig:images}
\end{figure*}

\section{High resolution imaging observations}
\label{sec:imaging}

\lens has been imaged at $\simeq 0.1$ arcsec FWHM resolution from the
optical (with \hst) to the NIR (with Keck Laser Guide Star Adaptive
Optics).  A summary of the imaging observations is given in
\Tref{tab:imaging}, whilst \Fref{fig:images} shows the {\it HST} and
Keck images of \lens. In this section we describe the multi-filter
high-resolution imaging data obtained for the \lens system in some
detail.

\begin{table}
 \centering
 \begin{minipage}{140mm}
  \caption{Summary of Imaging Observations}
  \begin{tabular}{cccc}
\hline  
\hline  
Telescope & Camera & Filter & Integration Time(s) \\
\hline
{\it HST} & WFPC2 & F450W & 4400 \\ 
{\it HST} & WFPC2 & F606W & 1600  \\
{\it HST} & ACS & F814W & 420 \\ 
Keck II & NIRC2-LGS & K' & 2700 \\
\hline
\hline 
\label{tab:imaging}
\end{tabular}
\end{minipage}
\end{table}


\subsection{ACS/WFPC2 imaging from \hst}
\label{sec:imaging:hst}

Hubble Space Telescope (\hst) observations of \lens were obtained on
June 12th 2006 with the Advanced Camera for Surveys (ACS), and on
April 19th 2009 with the Wide Field Planetary Camera 2 (WFPC2). The
ACS observation, in the \Ifilter (420s) filter, was part of the SLACS
snapshot programme GO:10587 (PI:Bolton); these data have a pixel scale
of $0.05$ arcsec. The WFPC2 observations, in the \Bfilter (4400s) and
\Vfilter (1600s) filters, were part of the cycle 16 supplementary
programme GO:11978 (PI:Treu). For the WFPC2 observations four
sub-exposures were obtained, and the frames drizzled to a pixel scale
of $0.05$ arcsec.

The \Ifilter image confirmed the strong lensing nature of this system
by showing that the background object was multiply imaged into a
three-component arc. It revealed that the lens was a disk dominated
galaxy, with a high inclination and dusty disk, and that the bulge was
compact and disk like.  The \Bfilter and \Vfilter images reveal that
the source is blue.

Models for the point spread functions (PSFs) of the ACS and WFPC2 data
where obtained by using the program TinyTim (Krist 1995). These PSF
models include the effects of sub-pixel dithering and drizzling and
have been found to provide adequate models for the true PSF (e.g.,
Bolton \etal 2008; Auger \etal 2009).


\begin{table*}
 \centering
 \begin{minipage}{1.0\linewidth}
   \caption{Summary of bulge plus disk fits together with stellar
     masses derived from SED fits with a Chabrier (2003) IMF.}
  \begin{tabular}{lccccccccc}
 \hline\hline
  & $q$ & $R_{50} [{\rm arcsec}]$ & $n$ & F450W-K' & F606W-K' & F814W-K' & K' magnitude & $\log_{10} (M_*/M_{\odot}) $ \\
\hline
Bulge & $0.53 \pm 0.02$ & $0.26 \pm 0.01$ & $1.21 \pm 0.11$ & $3.80 \pm 0.04$ & $2.40 \pm 0.03$ & $1.44 \pm 0.04$ & $17.76 \pm 0.28$ & $10.26\pm0.08$\\
 Disk & $0.31 \pm 0.02$ & $2.53 \pm 0.13$ & $\equiv1.0$ & $3.04 \pm 0.12$ & $1.75 \pm 0.14$ & $1.00 \pm 0.12$ & $16.25 \pm 0.13$ & $10.88\pm0.07$\\
  \hline
  \hline
\label{tab:brewfit}
\end{tabular}
\end{minipage}
\end{table*}

\subsection{NIRC2 imaging from Keck}
\label{sec:imaging:ao}

On August 13th 2009, we imaged \lens with the Laser Guide Star
Adaptive Optics (LGSAO) system on the Keck~II telescope. The tip-tilt
star had an R-band magnitude of~16.2 and a separation from the science
target of $60.4$ arcsec.

The images were taken in the \Kband with the near-infrared camera
NIRC2, in wide field format (with a $40 \times 40$ arcsec field of
view).  The pixel scale for this configuration is $0.04 \,\rm arcsec
\, pix^{-1}$. A total of 45 minutes of exposure was
obtained. Individual exposures were 1 minute in duration (divided into
two 30-second co-adds).  A dither was executed after every set of 5
exposures to improve sky sampling. Dithers were based on a four point
box pattern with sides 8~arcsec. The laser was positioned at the
center of each frame, rather than fixed on the central
galaxy. Observing conditions during the run were good.

The images were processed with the CATS reduction procedure described
by Melbourne \etal (2005).  A sky frame and a sky flat were created
from the individual science exposures after masking out all objects.
Frames were then flat-fielded and sky-subtracted.  The images were
de-warped to correct for known camera distortion. The frames were
aligned by centroiding on objects in the field, and finally co-added
to produce the final image.

A model for the PSF was derived from observations of a PSF star pair,
where the star used for tip-tilt correction is the same distance from
the PSF star as the lens galaxy was from its tip tilt star.  The star
pair observations were made immediately following the lens
observations.  The PSF star was found to have FHWM=$0.10$ arcsec (2.5
pixels) and a Strehl ratio of 18\%.

In the \Kband the extinction of both the lensed images and the lens
galaxy light due to dust in the lens galaxy is almost completely
absent, revealing a ring like structure, and confirming the disky
nature of the bulge. The background object appears to have been lensed
into a smooth arc in this filter.  The difference between the source
structure in the rest-frame NIR and the rest-frame UV/optical is
likely due to extinction from the lens galaxy artificially creating
the appearance of three distinct images.


\section{The stellar mass distribution}
\label{sec:mstar}

We begin our study of the mass distribution of \lens by inferring the
structure of the stellar component from the high resolution imaging
data described in the previous section. Following the standard
approach (e.g., MacArthur et al.\ 2003) our strategy is to model the
stellar mass distribution as an exponential disk of stars plus a
S\'ersic profile bulge, with each spatial component consisting of
distinct stellar populations. We first fit the surface brightness data
to obtain estimates of the shape and profile of the stellar mass
density, and then normalise the two profiles by fitting the bulge and
disk fluxes in our 4 filters (the spectral energy distribution, or
SED) with stellar population synthesis (SPS) models.


\subsection{Disk/bulge surface brightness fits}
\label{sec:mstar:bdfits}

In each band, a 2-dimensional model of the lens galaxy surface
brightness was fitted to the high resolution imaging data.  The model
is composed of two elliptically-symmetric \sersic~profile components,
representing the disk and the bulge.
\begin{equation}
\Sigma(x,y)=\Sigma_0\exp[-(R/R_0)^{1/n}]
\end{equation}
where $R = \sqrt{x^2 + y^2/q^2}$.  The S\'ersic index $n$ is fixed
at~1 for the exponential disk, and left free for the bulge.  The
remaining parameters for each component are the centroid position
$\{x_c,y_c\}$, scale radius $R_0$, the axis ratio $q$, and the
orientation angle~$\phi$. The prior probability distributions were all
independent, with uniform priors for $\phi$, $x_c$, $y_c$ and $q$, and
``Jeffreys'' ($\propto 1/x$) priors for $R_0$ and $n$, between
generous upper and lower bounds.

All four bands are fitted simultaneously, with all parameters except
for the normalization of the bulge and disk fluxes constrained to be
the same in all bands. This approach gives more robust colors of the
bulge and disk than is obtained when letting the structural parameters
float between bands.

The inferred parameter values for the disk and bulge surface
brightness are given in \Tref{tab:brewfit}.  For the bulge component,
we find a S\'ersic index of $\nb= 1.21 \pm 0.11$, and bulge
(luminosity) fraction which increases from $0.11 \pm 0.03$ in the
F450W filter to $0.20 \pm 0.05$ in the K' band. These values are
typical for low-redshift late-type spiral galaxies.

The bulge has a major-axis half-light radius of $\Reffb=0.26\pm 0.01''
= 0.63 \pm 0.02 \,\rm kpc$, whilst the disk has a major-axis
half-light radius of $\Reffd=2.53 \pm 0.13'' = 6.17 \pm 0.32 \,\rm
kpc$ corresponding to a disk scale length $R_{\rm d} = 1.51'' \pm
0.08'' = 3.68 \pm 0.19 \,\rm \kpc$.  The ratio between the bulge
half-light radius and the disk scale length is $0.17\pm0.02$ which is
consistent with those found by MacArthur \etal (2003) in a sample of
moderately inclined late-type spirals.

The bulge has an observed axis ratio of $\qb=0.53\pm 0.02$, while the
disk has an observed axis ratio of $\qd=0.31\pm 0.02$.  For a thin
disk the axis ratio equals the cosine of the inclination angle:
$\qd=\cos(i)$.  However, in general disks have a finite thickness,
which causes the true inclination to be higher than that inferred from
the observed axis ratio. For \lens we can infer the disk inclination
by measuring the axis ratio of the star forming ring, and assuming it
is intrinsically circular. This yields an axis ratio $q=0.20\pm0.02$,
and thus $i = 78.5 \pm 1.2$~degrees.

For an oblate ellipsoid with projected minor-to-major axis ratio, $q$,
and inclination, $i$, the 3D minor-to-major axis ratio, $q_3$ is given
by
\begin{equation}
q_3^2 = (q^2 - \cos^2i)/(1 - \cos^2i)
\end{equation}
Thus for \lens we infer that the 3D axis ratio of the bulge and disk
are $q_{3,\rm b}=0.51\pm0.02$, and ${q_{3,\rm d}=0.26\pm0.02}$. The
disk thickness that we derive for \lens is in good agreement with
measurements of edge-on spiral galaxies (Kregel \etal 2002).  We note
that the orientation angles of the disk and bulge components are very
close to each other: \lens appears to be well-modelled by an oblate
bulge bisected by a thick, coaxially symmetric disk.


\subsection{Stellar population SED fits}

The two component model for the surface brightness of \lens inferred
in the previous section can be used to constrain the stellar mass
distribution of the galaxy. Given the precision of the shape and
profile measurements, the normalisation of the stellar mass
distribution is the most uncertain part. Our aim is to constrain this
both gravitationally (via dynamical and lensing measurements), and by
modelling the stellar populations of the disk and bulge. In this
section we describe the latter route.

Assuming each spatial component has a distinct stellar population, we
can fit the model photometry for each component to SPS models and
infer the stellar mass of each component using the code describe in
Auger \etal (2009). We consider stellar populations characterised by
either a Chabrier (2003) or Salpeter (1955) IMF and described by 5
parameters: the total stellar mass~$\Mstar$, the population age~$A$,
the exponential star formation burst timescale $\tau$, the
metallicity~$Z$ and the reddening due to dust, $\tau_V$. We employ a
uniform prior requiring $9 \leq {\rm log_{10}} (\Mstar/\Msun) \leq 13$,
the age is constrained such that star formation began at some
(uniformly likely) time between $1 \leq z \leq 5$, $\tau$ has an
exponential prior with characteristic scale 1 Gyr, and we impose
uniform priors on the logarithms of the metallicity and dust
extinction such that $-4 \leq {\rm log_{10}} Z \leq -1.3$ and $-2 \leq
{\rm log_{10}} \tau_V \leq 0.3$. We note that the priors are the same
for the bulge and the disk components but are sufficiently
conservative that they do not bias our results. The posterior PDF is
sampled as described in Auger \etal (2009).

\begin{figure}
\centering
\includegraphics[width=0.46\textwidth,clip]
{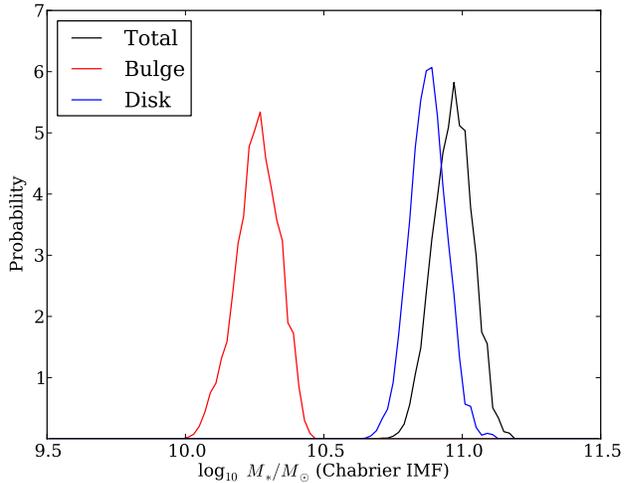}
\caption{Posterior distributions for the inference on the stellar mass
  based upon SPS models constrained by the high-resolution photometry,
  assuming a Chabrier IMF.}
\label{fig:mstarPosterior}
\end{figure}

\Fref{fig:mstarPosterior} shows the marginalised posterior inference
on the stellar mass, which we find to be well-constrained for the
bulge and disk components.  Assuming that both components are well
described by a Chabrier IMF, we find $\log_{10}(\Mstarb/\Msun) = 10.26
\pm 0.08$ and $\log_{10}(\Mstard/\Msun) = 10.88 \pm 0.07$, for the
bulge and disk respectively, justifying our description of \lens as
``disk-dominated.''  The total stellar mass of \lens from SED fitting
is therefore $\log_{10}(\Mstar/\Msun) = 10.97 \pm 0.07$, and the bulge
fraction is $f_{\rm bulge}=0.20\pm 0.04$ (the same as in the K'-band
light). For a Salpeter IMF the masses are all 0.24 dex higher. We will
return to this inference in \Sref{sec:imf} below, where we compare it
to the stellar mass implied by the gravitational analysis.


\begin{table*}
\centering
\begin{minipage}{0.60\linewidth}
  \caption{Summary of mass model priors. Bold indicates an
    uninformative prior, regular text indicates an informative prior
    (i.e. that the parameter is virtually
    fixed). $L\mathcal{N}(a,b^2)$ denotes a lognormal distribution,
    with $a$ being the central value for the variable, and $b$ being
    the standard deviation for the log of the
    variable. $\mathcal{N}(a,b^2)$ denotes a normal distribution, with
    $a$ being the central value and $b$ being the standard deviation.
    $U(a,b)$ denotes a uniform distribution with lower and upper
    limits, $a$ and $b$, respectively. For clarity we have arranged
    the parameters into three groups: free parameters first (with
    uninformative priors), stellar mass distribution parameters (with
    priors derived from the SPS modelling of \Sref{sec:mstar}), and
    finally nuisance parameters. }
\label{tab:priors}
\begin{tabular}{llll}
  \hline
  \hline
  parameter                 & description                             & prior \\
  \hline
  $\Vhalo/\kms$             & dark halo asymptotic circular velocity  & {\bf $\mathcal{N}$(280, 50$^2$)}\\              
  $\qhalo$                  & dark halo 3D axis ratio         & {\bf L$\mathcal{N}$(1, 0.3$^2$)}\\
  $\rhalo$/arcsec           & dark halo core radius                   & {\bf U(0.01, 10)} \\
  $\logMstar$               & stellar mass                            & {\bf U(10.5,11.4)} \\  
  \hline
  $f_{\rm bulge}$            & bulge stellar mass fraction             & $\mathcal{N}(0.2, 0.04^2)$ \\ 
  $q_{\rm bulge}$            & bulge 2D axis ratio                     & $L\mathcal{N}(0.53, 0.03^2)$  \\
  $R_{0, {\rm bulge}}$/arcsec & bulge chameleon size                    & $L\mathcal{N}(0.094, 0.03^2)$ \\
  $\alpha_{\rm bulge}$       & bulge chameleon index                   & 0.4892 \\
  $q_{\rm disk}$             & disk 2D axis ratio                      & $L\mathcal{N}(0.31, 0.03^2)$ \\
  $R_{0, {\rm disk}}$/arcsec  & disk chameleon size                     & $L\mathcal{N}(1.10, 0.03^2)$\\ 
  $\alpha_{\rm disk}$        & disk chameleon index                    & 0.63 \\
  $\cos(i)$                & cosine of disk inclination angle        & 0.2 \\ 
  \hline
  $x_{\rm c}$/arcsec        & spatial offset in x direction            & $\mathcal{N}(0, 0.01^2)$ \\ 
  $y_{\rm c}/$arcsec        & spatial offset in y direction            & $\mathcal{N}(0, 0.03^2)$ \\ 
  $\theta$/deg             & mass-light position angle offset         & $\mathcal{N}(1.7, 2.9^2)$ \\ 
  $\gamma$                 & lens external shear                      & $\mathcal{N}(0, 0.1^2)$ \\
  $\theta_\gamma$/deg       & position angle of external shear         & $U(0, 180)$ \\ 
  \hline
  \hline
\end{tabular}
\end{minipage}
\end{table*}

\section{A three-component galaxy mass model}
\label{sec:model}

As indicated in the previous section, it makes sense to consider a
stellar mass distribution for \lens whose profile and spatial shape is
tightly constrained by the structural analysis of the surface
brightness observed. However, we would like to remain agnostic about
the stellar IMF, and instead investigate the stellar mass of the
galaxy independently, using strong lensing and stellar
kinematics. Since these are sensitive to the total mass of the galaxy,
we include a dark matter halo component to the model as well.  In this
section we describe this 3-component mass model in some detail,
including the predictions it makes for the observable effects, and the
prior PDFs we assign on the model parameters.


\subsection{Description}
\label{sec:model:descr}

Based on the results of the previous section, we model the stellar
mass distribution of \lens as a thin, circular exponential disk of
stars of mass $\Mstard$, co-axial with an oblate bulge of stars of
mass $\Mstarb$.  We then assume the galaxy to reside in a dark matter
halo that is also axisymmetric, and aligned and concentric with the
disk and bulge. This assumption that the galaxy and inner dark matter
haloes are aligned is supported by cosmological simulations of disk
galaxy formation (e.g., Deason et al. 2011). We note that for our
strong lensing analysis it is feasible to allow the position angles of
the baryons and dark matter to be offset. However, this would make the
model non-axisymmetric and thus make the kinematics considerably
harder to model. The surface brightness in our four filters constrains
the spatial distribution of stellar mass tightly, under the assumption
that the stellar mass-to-light ratios are radially constant; we leave
the overall normalisation of the stellar mass distribution as a free
parameter.

We do not explicitly include a cold gas disk for two reasons. Firstly,
we do not have direct observations of the atomic and molecular gas in
\lens. Secondly, observations suggest that the gas fractions for
low-redshift luminous spiral galaxies are of order $20\%$ (e.g.,
Dutton \& van den Bosch 2009), and thus do not contribute
significantly to the baryonic mass.  However, if we were to assume
that any cold gas present has the same spatial distribution as the
stars, the unknown gas mass could be absorbed into a baryonic
mass-to-light ratio that includes the stellar mass-to-light ratio. For
the majority of this paper we neglect the cold gas mass, but return to it
in the discussion of the IMF (\Sref{sec:imf}) below.

The three mass components are described as follows:
\begin{itemize}
\item Exponential Stellar Disk
\item S\'ersic Stellar Bulge
\item Non-singular Isothermal Ellipsoid (NIE) Dark Matter Halo
\end{itemize}
This model has 17 parameters in total; they, and their prior PDFs, are
given in \Tref{tab:priors}. We assign informative priors to all but 4
of these parameters, propagating the uncertainties in the surface
brightness fits through to the mass model.

First, we assume the bulge, disk, and halo inclination are all the
same, and given by the thin disk axis ratio (0.2), as in
\Sref{sec:mstar} -- we assume that this is known with no uncertainty.
As we describe below, we use an approximation to the Exponential
profile that allows us to compute predicted observable quantities
efficiently -- the size parameters of the bulge and disk in that
approximation are determined from the results of the previous section,
as is (more straightforwardly) the bulge axis ratio. We use the
derived value of 1.21 (see Table \ref{tab:brewfit}) for the Bulge
S\'ersic index, with no uncertainty.

We assume that the disk and bulge are different stellar populations,
and so use the independent stellar mass results from the previous
section to constrain the bulge mass fraction, $f_{\rm
  bulge}=\Mstarb/\Mstar$. As already mentioned, we leave the total
stellar mass $\Mstar$ as a free parameter with uniform prior on its
logarithm. This is effectively equivalent to assuming that the two
components have very similar, although unknown, IMF normalization, We
do inform the bounds of this uniform prior using the SPS modelling
results, in the following way. Estimating that the lightest
conceivable IMF would give stellar masses systematically a factor of
two lower than Chabrier, we take the 3-sigma point of the Chabrier PDF
in \Fref{fig:mstarPosterior} and subtract 0.3~dex to set a lower limit
on $\logMstar$ of 10.5. Likewise, at the high end we take Salpeter to
be the heaviest IMF and use the 3-sigma point of the Salpeter PDF in
\Fref{fig:mstarPosterior} (11.4) as our upper limit on $\logMstar$.
We note that none of our results change if we adopt a higher upper
limit to the stellar mass.

This leaves 3 parameters that describe the model dark matter halo:
$\Vhalo$, $\rhalo$, and $\qhalo$.  We allow the axis ratio of the halo
to be greater than unity, corresponding to a prolate halo, but use a
broad lognormal distribution centred on spherical to encode
approximately our expectations. For the halo density profile, the NIE
profile has considerable freedom and can represent a much broader
range of behaviours than those seen in simulation. Therefore, we adopt
physically motivated priors to select the cosmologically motivated
subset of parameters combination. Studies of large sets of spiral
galaxies, using satellite kinematics and weak galaxy-galaxy lensing,
in the context of numerical simulations have shown that the maximum
observed circular velocity is typically comparable to the maximum
circular velocity of the halo, even though these two maximums occur at
vastly different radii (Dutton \etal 2010a). We also know that rotation
curves do not keep rising indefinitely, but typically flatten out
within a few scale radii of the disk. To inject this information we
require that the asymptotic circular velocity of the halo be
comparable to that measured via spectroscopy (see \S~\ref{sec:spec}
below): in practice we assign a broad Gaussian prior centred on
280~\kms with width 50 \kms. The prior is chosen to be broad enough --
the 3-sigma range of this Gaussian spans the range 130 to 430~\kms --
not to drive the final inference and yet tight enough to rule out
models where the maximum velocity is reached too far out. In addition
we impose a uniform prior PDF for the core radius, allowing it to be
at most 24~kpc (10'').

In later sections we will introduce the kinematic and lensing data,
and then use them to constrain the parameters of this 3-component mass
model. However, before getting to the data, in the rest of this
section we give the functional forms for each mass component, and the
predicted observables resulting from them.


\subsection{Three-dimensional component density profiles}
\label{sec:model:profiles}

The axisymmetric ellipsoidal halo is assumed to have a non-singular
isothermal (NIE) profile, which we parametrise in a cylindrical
coordinate system in the plane of the galaxy following Keeton \&
Kochanek (1998):
\begin{equation}
\label{eq:chm3d}
  \rho_{\rm NIE}(R,z;\vc,\rc,\q3) = 
     \frac{\vc^2}{4 \pi G \q3} 
     \frac{e}{\sin^{-1}e} 
     \frac{1}{\rc^2 + R^2 +z^2/\q3^2}.
\end{equation}
Here, $\vc$ is the asymptotic circular velocity, $\rc$ is the core
radius, $\q3$ is the three dimensional axis ratio, and
$e=(1-\q3^2)^{1/2}$ is the eccentricity. For a zero thickness mass
distribution ($\q3=0$), $e/\sin^{-1}e = 2/\pi$. For a spherical mass
distribution ($\q3=1$), $e/\sin^{-1}e = 1$. For a prolate mass
distribution ($\q3>1$), while $e$ is imaginary, $e/\sin^{-1}e$ is real
and greater than 1.

This mass profile is often used in gravitational lens analysis, since
its projected mass distribution and deflection angles can be computed
analytically (Keeton \& Kochanek 1998).  This model has been used very
successfully to model the total (dark plus stellar) mass profiles of
elliptical galaxy lenses (e.g., Bolton \etal 2008). In this work we
use the NIE model for the halo alone. While the NIE profile has a
constant central density, it is flexible enough to broadly capture the
change in the density profile in the central regions that we expect
from numerical simulations of dark matter halos (e.g., Navarro \etal
1997).

We would like to model the stellar disk and stellar bulge mass
components such that in projection they appear to have exponential and
S\'ersic profiles respectively. However, we also need 3D distributions
for which we can compute predicted rotation curves, as well as
projected distributions convenient for lensing calculations.  To
achieve this we note that the NIE profile can be used to create an
approximation to an exponential profile in projection (Maller \etal
2000).  This is done by taking the {\it difference of two NIEs}.  If
$\rho_{\rm NIE}(R, z; \vc,\rc,\q3)$ is a softened isothermal
ellipsoid, then
\begin{eqnarray}
  \rho_{\rm Chm}(R, z; \vc, \rc, \q3, \alpha) &=& 
     \rho_{\rm NIE}(R, z; \vc, \rc, \q3) \\ \nonumber
 &-& \rho_{\rm NIE}(R, z; \vc, \rc/\alpha, \q3)
\end{eqnarray}
is a ``Chameleon'' profile with positive density everywhere, and a
finite total mass.  In Appendix~\ref{app:chameleon} we derive new
formulae that provide Chameleon approximations to S\'ersic profiles of
any index (for $1 \lta n \lta 4$), providing the Chameleon size $\rc$
and index $\alpha$ given a S\'ersic half light radius $\Reff$ and
index~$n$.


\subsection{Predicted rotation curves}
\label{sec:model:rotcurves}

For our ellipsoidal mass profiles, we can calculate the rotation
velocity, as a function of radius, of a massless test particle moving
on a circular orbit in the plane of the galaxy. We refer to this
velocity as the circular velocity to distinguish it from the rotation
velocity of the stars and gas, which may be lower than the circular
velocity due to a velocity dispersion component.  The circular
velocity profile for the NIE model is (Keeton \& Kochanek 1998)
\begin{eqnarray}
\label{eq:vnie}
\frac{V_{\rm NIE}^2(R; \vc,\rc,\q3)}{\vc^2} = %
  1 - \frac{e}{\sin^{-1}e} \frac{\rc}{(R^2 +e^2\rc^2)^{1/2}} \nonumber \\
\times \tan^{-1}\left[ \frac{(R^2 + e^2 \rc^2)^{1/2}}{\q3\rc} \right],
\end{eqnarray}
where again $e=(1-\q3^2)^{1/2}$ is the eccentricity of the mass
distribution and the model is normalised so that, asymptotically for
$R\rightarrow \infty$, $V_{\rm NIE}(R) \rightarrow V_c$.
For the special case of a zero thickness mass distribution
($\q3=0,e=1$) \Eref{eq:vnie} reduces to
\begin{equation}
\label{eq:vnieq0}
  \frac{V_{\rm NIE}^2(R)}{\vc^2} = 1 - \frac{\rc}{(R^2 +\rc^2)^{1/2}}.
\end{equation}
For the case of a prolate mass distribution ($\q3 > 1$),
$e/\sin^{-1}e$ is real, but since $e^2 < 0$, for $(R^2 +e^2\rc^2) < 0$
the circular velocity is given by
\begin{eqnarray}
\label{eq:vniepro}
  \frac{V_{\rm NIE}^2(R)}{\vc^2} = %
  1 - \frac{\tilde{e}}{\sinh^{-1}\tilde{e}} %
  \frac{\rc}{(-R^2 -e^2\rc^2)^{1/2}} \nonumber \\
  \times \tanh^{-1}\left[ \frac{(-R^2 - e^2 \rc^2)^{1/2}}{\q3 \rc} \right],
\end{eqnarray}
where $\tilde{e}=\sqrt{|e^2|}$.

For the chameleon profile the circular velocity is given by the
quadratic difference between the circular velocities of the
sub-component NIE's:
\begin{eqnarray}
  V^2_{\rm chm}(R; \vc,\rc,\q3,\alpha) = V^2_{\rm NIE}(R; \vc,\rc,\q3) \\ \nonumber
  - V^2_{\rm NIE}(R; \vc,\rc/\alpha,\q3).
\end{eqnarray}
Likewise, for the mass model the total circular velocity is given by
the quadratic sum of the circular velocities of the bulge, disk, and
halo components:
\begin{equation}
  V^2(R) = V^2_{\rm bulge}(R) +V^2_{\rm disk}(R) +V^2_{\rm halo}(R)
\end{equation}.


\subsection{Predicted lensed images}
\label{sec:model:deflection}

Projecting the three components onto the sky allows us to compute
deflection angles and predict the observed gravitational arc, pixel by
pixel.

In projection the mass distribution (an oblate or prolate ellipsoid
with minor to major axis ratio $q_3$) has projected axis ratio $q$
given by
\begin{equation}
  q=(q_3^2\sin^2 i +\cos^2 i)^{1/2}, 
\end{equation}
where $i$ is the inclination angle (such that $i=0^{\circ}$
corresponds to a face-on disk, and $i=90^{\circ}$ to an edge-on one).
In general the projected axis ratio, $q$, will be
closer to unity than the 3D axis ratio, $q_3$. 

The projected mass density of an NIE model is given by (Keeton \&
Kochanek 1998):
\begin{eqnarray}
\label{eq:nie2d}
  \Sigma_{\rm NIE}(x,y;b,\rc,q) &=& \frac{\vc^2}{4 G \Dd} 
      \frac{e}{\sin^{-1}e}
      \frac{1}{q\sqrt{r_c^2 + x^2 + (y/q)^2}}, \nonumber \\
                                &=& \frac{\Sigma_{\rm crit}b}{2}
      \frac{1}{q\sqrt{r_c^2 + x^2 + (y/q)^2}}.
\end{eqnarray}
Here, $\Dd$ is the angular diameter distance to the lens, and $e$ is
again the ellipticity, while in the second line $b$ is the minor axis
of the critical curve (and thus $b/q$ is the major axis of the
critical curve), and $\Sigma_{\rm crit}$ is the critical surface
density of strong lensing:
\begin{equation}
\Sigmacrit = \frac{c^2}{4 \pi G}\frac{\Ds}{\Dds \Dd},
\end{equation}
where $\Ds$ is the angular diameter distance from the observer to the
source, and $\Dds$ is the angular diameter distance from the lens to
the source. For our assumed cosmology, for \lens these distances are:
$\Dd=497.6 \rm Mpc$, $\Ds=1510.2 \rm Mpc$, $\Dds=1179.6 \rm Mpc$, and
thus the critical density is $\Sigmacrit=4285.3 \rm M_{\odot}
pc^{-2}$.

To explain the parts of \Eref{eq:nie2d} a little further, the
parameter, $b$, is related to the spherical Einstein radius, $\bsis$,
via:
 \begin{equation}
   b=\bsis (e/\sin^{-1}e),
 \end{equation}
 and the spherical Einstein radius (in radians) is in turn related to
 the asymptotic circular velocity, $\vc$, via:
 \begin{equation}
\label{eq:bsis}
 b_{\rm SIS} = 2\pi (\vc/c)^2 \Dds/\Ds. 
 \end{equation}

 The deflection angles are given by (Keeton \& Kochanek 1998)
\begin{eqnarray}
\alpha_x &=& \frac{b}{ (1-q^2)^{1/2}} \tan^{-1}\left[ \frac{(1-q^2)^{1/2}x}{\Psi+\rc}\right], \\
\alpha_y &=& \frac{b}{ (1-q^2)^{1/2}} \tanh^{-1}\left[ \frac{(1-q^2)^{1/2}y}{\Psi+q^2\rc}\right], \\
\end{eqnarray}
where $\Psi^2 = q^2(\rc^2+x^2)+y^2$.  The deflection angles from the
three components of the mass distribution can be simply summed, as
they are just the first derivatives of the projected (lens) potential
of each component, and the potentials of the three components can be
summed themselves to give the total potential. Likewise, the Chameleon
profile in projection is just the difference between two projected NIE
models, and its deflection angles are just the difference between
those of its NIE components.

To predict the positions and structure of the lensed images given a
set of mass model parameters, we map each observed pixel location back
to the source plane using the overall deflection angle map, and look
up the surface brightness of a model source at that position. In
practice we use a single, elliptically symmetric source with a
S\'ersic brightness profile, as in e.g., Marshall \etal (2007).


\section{Strong Gravitational Lensing Data}
\label{sec:lens}

We now present the strong gravitational lensing data that we will use
to constrain our mass model. We first describe the preparation of the
arc imaging data, and then show with a simple lens model the
information it contains.


\begin{figure}
\centerline{
\psfig{figure=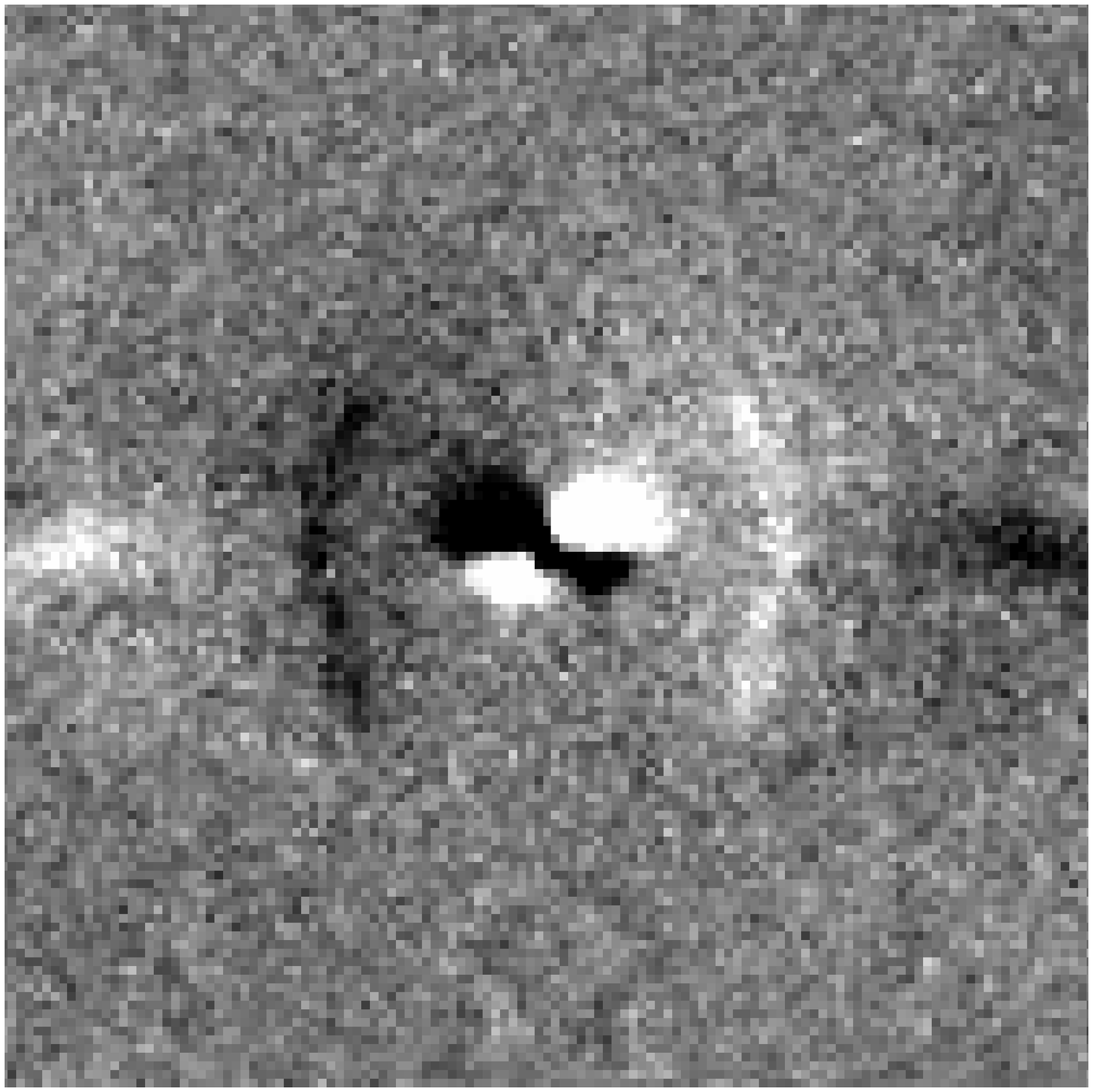,width=0.40\textwidth}
}
\vspace{0.5cm}
\centerline{
\psfig{figure=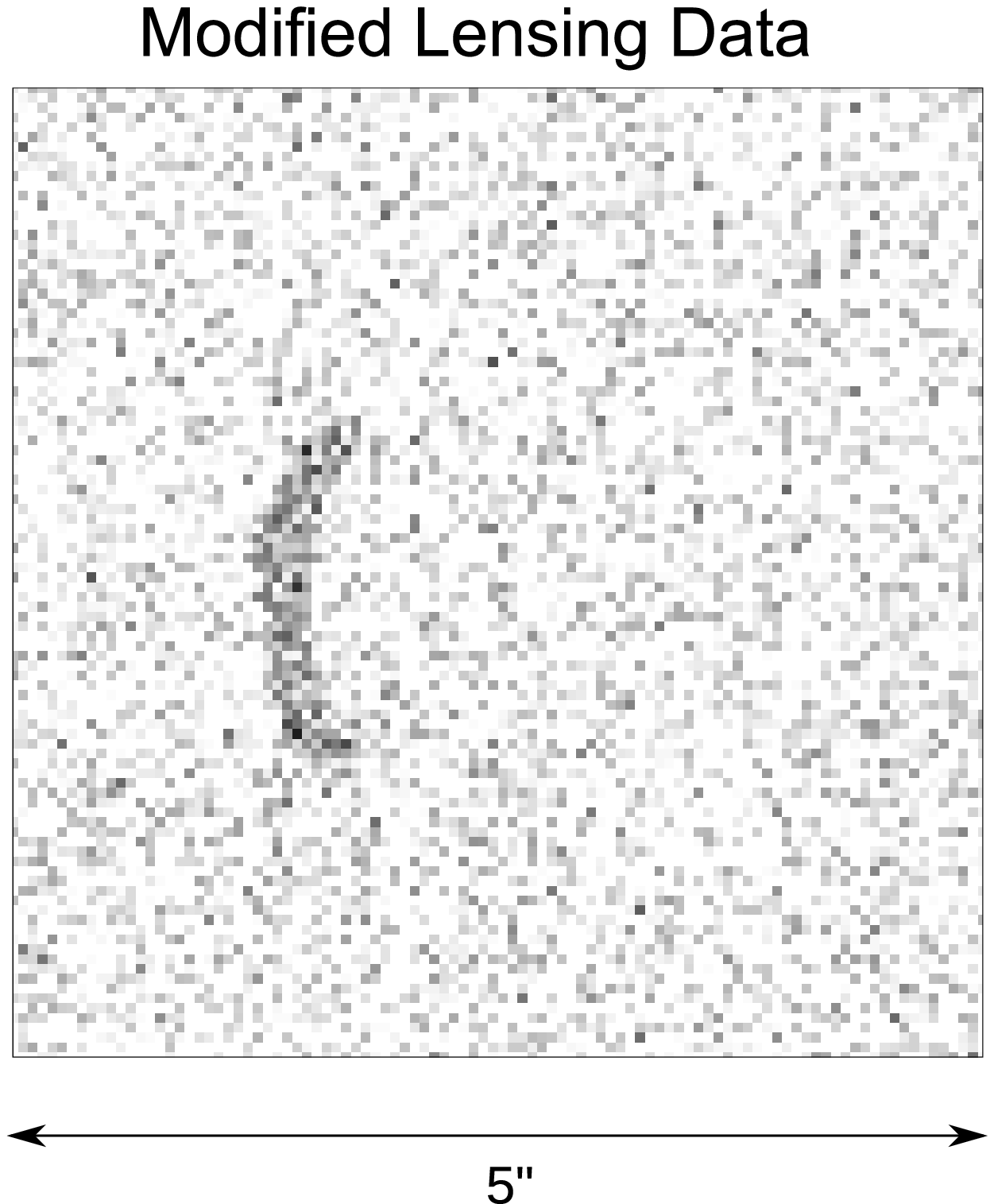,width=0.40\textwidth}
}
\caption{Upper panel: Galaxy subtracted image in the \Kband obtained using 
a reflection of the image. The lens galaxy is well subtracted near the arc, 
but there are significant residuals near the center. Lower panel: An arc 
in the right place, obtained by masking out non-arc
features from the image in the upper panel. This was used for the actual
fitting, to weaken the fit criterion. Essentially, we want
to produce models whose posterior distribution predicts an arc
with this morphology, but does not necessarily need to match every pixel.
\label{fig:just_arc}}
\end{figure}

\subsection{The lensed arc likelihood}
\label{sec:results:lhood}

Due to the strong effects of dust in the lens system, we focus our
lensing analysis primarily on the \Kband NIRC2 image.  In the \Kband
the lens galaxy appears to be much smoother than in the optical, but
the light distribution is not able to be modelled by a simple surface
brightness profile. This makes subtracting the lens galaxy light
difficult. Our goal is to obtain robust parameter inferences with
meaningful uncertainties, and so we opt for quite a conservative
application of the imaging data. To account for lens subtraction
errors, we create an arc image and a goodness of fit statistic that
rewards a model for having an arc in the right place, with the right
shape, and that is all: we do not require the detailed features of the
modified surface brightness profile to be matched.

To achieve this, we first subtracted the galaxy light around the arc
by reflecting the galaxy along the minor axis. This method provides a
better subtraction than multiple S\'ersic components, or a radial
bspline model (as used by e.g., Bolton \etal 2006).  We then cut out
the arc and set the remaining pixels to zero, before adding noise at
the level of $\sigma_{15} = 15$\% of the peak arc brightness.  This
15\% value is an initial estimate of the appropriate noise level
needed to suppress models that predict significant lensed features
elsewhere, although faint counter-images are still allowed.  The
resulting modified image is shown in \Fref{fig:just_arc} (lower
panel); we also show the lens subtracted image (upper panel), with its
uncertain central region. It is not clear whether or not there is a
counter-image in the centre of the system. We note that if there is no
counter image, then \lens would be a rare example of galaxy-scale
naked cusp lens.

The likelihood function for the modified image data was that used by
Brewer \& Lewis (2006), Marshall \etal (2007), and others. We assume
Gaussian errors of $\sigma_{15}$ on the pixel values $\data$; we can
predict these pixel values from a model source with parameters
$\srcpars$ (as described in \Sref{sec:model:deflection} above) given
lens model parameters $\masspars$. Denoting the predicted data as
$\datap$, we write down the usual chi-squared misfit function
\begin{equation}
  \chi^2 = \sum_i^{\rm pixels} %
  \frac{\left[\datai -\datapi(\masspars,\srcpars)\right]^2} {2\sigma_{15}^2}.
\end{equation}
We allow the data to inform our understanding of the model
uncertainty, by re-scaling the denominator by a factor $T$. This
corresponds to increasing or decreasing the perceived errors on the
pixel values, and provides a mechanism for avoiding over-fitting the
arc structure or allowing models that predict undetected flux. (The
symbol $T$ stands for ``temperature'' -- increasing the temperature
increases the diffusion of the model around its parameter space.)  The
likelihood function is then:
\begin{equation}
  \pr(\data|\masspars,\srcpars) \propto \exp\left(-\frac{1}{2T}\chi^2\right)
\end{equation}

The value of $T$ selected was 7.5. This was the highest value where
the posterior distribution only contained images that resemble the arc
morphology. Higher temperatures caused the posterior models to predict
substantial flux that is not observed, lower temperatures enforced the
fit to the modified image to be too strict. The reason for the
two-step procedure (adding noise and then selecting a temperature) is
that Nested Sampling provides the results for all temperatures in a
single run, whereas tuning the noise level itself would have required
large numbers of trial runs.

Using this modified image, and the temperature-raising scheme, allows
us to explore an approximate posterior distribution for the lens model
parameters that conditions on the presence of an arc with the observed
morphology, and nothing else. 

\begin{figure}
  \centering\includegraphics[width=0.95\linewidth]{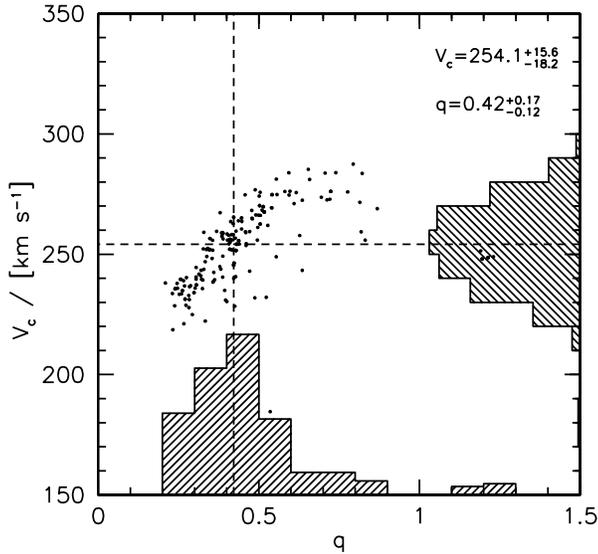}
  \caption{Marginalised posterior PDF for the 2D axis ratio, $q$, and
    circular velocity, $V_{\rm c}$, for a single SIE component model
    fitted to the strong lensing data. The dashed lines show the
    median values of $q$ and $V_{\rm c}$.}
  \label{fig:nielensfit}
\end{figure}

\subsection{Constraints on projected ellipticity and mass}
\label{sec:lensq}

To illustrate the unique information that strong gravitational lensing
provides, we first perform a fit to the lensing data with a single
singular isothermal ellipsoid (SIE) mass model. The purpose of this
exercise is to show that strong lensing places constraints on the axis
ratio of the projected mass, as well as the projected mass within the
Einstein radius.

Fixing its centroid and orientation to that of the lens surface
brightness, our example SIE lens model has two parameters, minor axis
Einstein radius $b$, and axis ratio $q$. We assign uniform priors over
wide ranges for these parameters, and then explore the posterior PDF
using our sampling code (which we introduce in more detail in
\Sref{sec:results} below).  We find the circularised Einstein radius
to be $\theta_{\rm Ein} = b/\sqrt{q} = 0.89^{+0.05}_{-0.08}$~arcsec,
and the axis ratio to be $q = 0.42^{+0.17}_{-0.12}$.  We can transform
samples in $b$ and $q$ into the circular velocity, which we expect to
be well constrained. \Fref{fig:nielensfit} shows the marginalised
posterior PDF for $q$ and $V_{\rm c}$ -- the circular velocity is
indeed well-constrained: $V_{\rm c}=254^{+15}_{-18}\kms$. The shape of
the arc also constrains the ellipticity of the total mass
distribution: since in our three-component model the ellipticity of
the disk and bulge are fixed, we expect strong lensing to then provide
information about the shape of the dark halo.

\begin{figure*}
\centering\includegraphics[width=0.99\linewidth]
{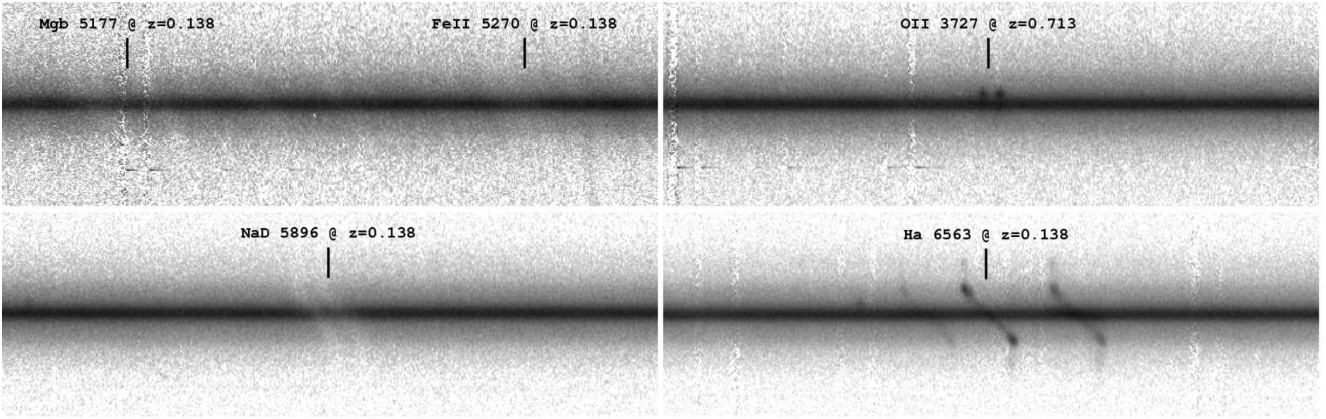}
\caption{Cutout images of the DEIMOS optical long slit spectrum of
  \lens centered on prominent emission and absorption features: Mgb
  5177, FeII 5270 from the lens (upper left); NaD 5896 from the lens
  (lower left); OII 3727 from the source (upper right); and H$\alpha$
  6563, [NII] 6550 6585 from the lens (lower right).  The vertical
  scale is 20 arcsec; the vertical bar is 3 arcsec. Up corresponds to
  East (left in the images in \Fref{fig:images}). The ring in the
  imaging (Fig.~\ref{fig:images}) corresponds to strong H$\alpha$
  emission due to star formation.}
\label{fig:spec}
\end{figure*}


\section{Gas and Stellar Kinematics Data}
\label{sec:spec}

The second set of data that we will use to constrain our mass model is
a galaxy rotation curve, derived from optical emission and absorption
line spectroscopy. In this section we describe the observations, and
the rotation curve extraction process, discuss the observed velocity
dispersion and our interpretation of it, and then derive the
likelihood function that we will use when fitting our mass model.

\subsection{Spectroscopic observations with Keck}

Major axis ($\rm PA=87^{\circ}$) long-slit spectra were obtained with
the DEep Imaging Multi-Object Spectrograph (DEIMOS), and Low
Resolution Imaging Spectrograph (LRIS) on the Keck 10-m telescopes.

On October 1st 2008 \lens was observed with DEIMOS on Keck II. We used
the 1200 line grating (corresponding to a pixel scale
of~$0.32\Angstrom$) with a 1'' width slit resulting in a spectral
resolution of $\simeq 1.9 \Angstrom$. The central wavelength was
$6500\Angstrom$, resulting in a wavelength range of
$5200-7800\Angstrom$. We took three exposures of 1200s in seeing
conditions of $0.60''$. The slit was aligned with the major axis of
the galaxy, with $\rm PA=87$. The spectra were reduced using routines
developed by D.~Kelson (Kelson 2003).

On November 27th 2008 we observed \lens with LRIS on Keck I.  However,
the seeing for this observation was considerably worse ($\simeq
1.5''$) than that of our DEIMOS observation. This resulted in
increased beam-smearing and reduced sensitivity, and thus we focus our
kinematic analysis on the DEIMOS observations.


\subsection{The observed rotation curve}
\label{sec:spec:rotcurve}

Cutouts of the DEIMOS long-slit spectrum centered around prominent
emission and absorption lines are shown in \Fref{fig:spec}.  These
show clear signs of rotation in both emission and absorption
lines. For the emission lines we measured rotation curves by locally
fitting Gaussian line profiles to one-dimensional spectra extracted
along the slit. For the absorption lines we measured the rotation and
dispersion profile by applying \python routines developed by M.W.Auger
to one-dimensional spectra extracted along the slit.  The upper right
panel also shows the spatially offset \OII emission lines from the
source galaxy.

\begin{figure}
\centering\includegraphics[width=0.99\linewidth]
{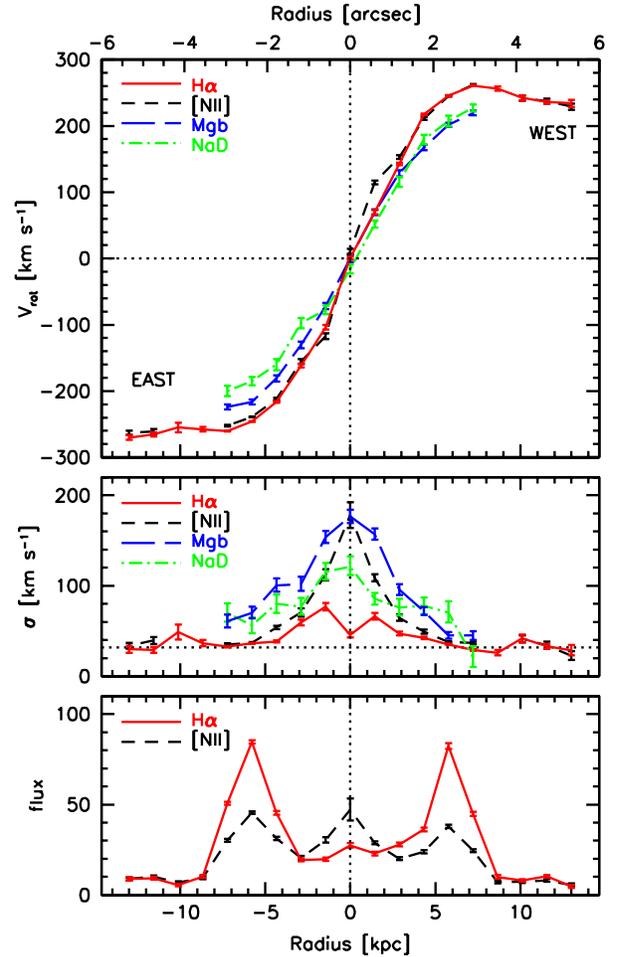}
\caption{Rotation curve (upper panel), velocity dispersion profile
  (middle panel), and line flux profile for \lens. }
\label{fig:rc}
\end{figure}

The extracted rotation curve is shown in the upper panels of
\Fref{fig:rc}. The spatial sampling is $\simeq 0.59"$ (5 DEIMOS
pixels), corresponding to 1 data point per seeing FWHM.  There is good
agreement in the rotation curves measured in \Ha and \NII, except near
the very center, where \NII gives a higher $V_{\rm rot}$. This is
possibly due to \Ha being contaminated with stellar absorption. The
\NaD and \Mgb absorption lines give lower $V_{\rm rot}$ than the
emission lines, especially at larger radii. This is expected due to
the increased pressure support in the stars compared to the gas, so
called asymmetric drift.

The rotation curve flattens out beyond 3~arcsec (7~kpc), corresponding
to 2 disk scale lengths. On the East side the rotation curve remains
flat out to the last data point (13~kpc). On the West side the rotation
curve decreases beyond 3.5~arcsec (8.5~kpc).  We trace this asymmetry
to the warp in the West side of the optical disk, which causes the
slit to miss the major axis: beyond 3.5~arcsec we therefore use only
the East side of the rotation curve.

\Fref{fig:rcfold} shows the folded rotation curve obtained by
combining the rotation curves from \Ha and \NII. The data points shown
in this figure are given in \Tref{tab:rc}. When combining data points
we use the error weighted mean. The new error is the maximum of the
statistical error and half the differences between the two data
points.  From this rotation curve the maximum observed rotation
velocity is $271\pm 4 \kms$ (corrected for inclination, but not beam
smearing) at 13 kpc from the galaxy center.

\subsection{The rotation curve likelihood}
\label{sec:spec:lhood}

For a given set of mass model parameters~$\masspars$ we can predict
the circular velocity of the stars at each radius, and hence find
parameter vectors that fit the rotation curve data.  To do this we
need to fold in the effects of beam-smearing; our procedure for this
is described in the next section.  We compare the predicted data
$\vrotp$ and the observed data in the usual way.  We assume
uncorrelated Gaussian errors on the observed velocities $\vrotj$
$\sigma_j = \sqrt{K \sigma_{0, j} + \sigma_{\rm extra}^2}$ (where
$\sigma_{0, j}$ are the reported error bars, and $K$ and $\sigma_{\rm
  extra}$ are free parameters allowing the true uncertainties to be
inferred), and hence construct the familiar likelihood function
\begin{equation}
  \pr(\vrot|\masspars) = \frac{\exp{\left(-\frac{\chi_v^2}{2}\right)}}
                           {\prod_{j=1}^n \sigma_j \sqrt{2\pi}}
\end{equation}
where the misfit function
\begin{equation}
  \chi_v^2 = \sum_j \frac{\left(\vrotj - \vrotpj\right)^2}{\sigma_j^2}.
\end{equation}
We then take the product of this likelihood and the prior PDFs on the
parameters defined in \Sref{sec:model} to obtain the posterior PDF for
the model parameters given the kinematics data.


\subsection{Modelling beam-smearing}

\begin{figure}
\centering\includegraphics[width=0.99\linewidth]
{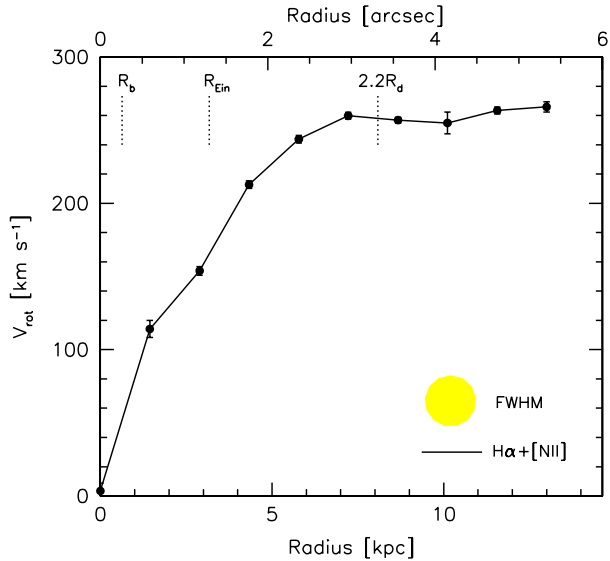}
\caption{Folded rotation curve from emission lines of H$\alpha$ and
  [NII]. The location of the bulge half-light radius, $R_{\rm b}$,
  Einstein radius, $R_{\rm Ein}$, and 2.2 disk scale lengths, $R_{\rm
    d}$ are indicated with dotted lines. The spatial sampling is one
  point per seeing FWHM, which is indicated by the yellow circle. This
  rotation curve is uncorrected for inclination, and beam smearing due
  to the finite (1 arcsec) slit width and seeing.}
\label{fig:rcfold}
\end{figure}

The rotation curve data presented in \Fref{fig:rcfold} are the
observed values, uncorrected for inclination, finite slit width and
seeing effects. We refer to these combined effects as
beam-smearing. Since the disk inclination is high, the $\sin i$
correction is small, just a factor of $1.021$. However, since the slit
width covers a large fraction of the minor axis of the galaxy, the
effects of finite slit width and seeing are likely to be significant,
especially near the centre of the galaxy.  We take this into account
when computing the predicted data $\vrotp$ as the inclined,
beam-smeared, model rotation curve within a 1~arcsec slit.  For
computational efficiency we estimate the beam-smearing effect using a
simplified, rotating exponential disk model, and then apply this
correction to the model rotation curve. The intrinsic rotation curve
$\vrotmodel$ is given by the sum of the bulge, disk and halo
components as described above.

The beam-smearing calculation is approximate, because we don't know
the exact distribution of the \Ha emission, only the starlight.  While
we model the \Ha distribution with an exponential profile, with the
scale length of the V-band light, the actual distribution is likely to
be asymmetric (due to extinction), and non-exponential (there is a
ring of star formation). To minimize the impact of these
uncertainties, we have excluded the inner $2$ arcsec of data in our
mass models.

\begin{table}
 \centering
 \begin{minipage}{140mm}
  \caption{Observed Rotation Curve from Emission Lines}
  \begin{tabular}{cccc}
\hline  
\hline  
Radius & Radius & Rotation Velocity & Error \\
 $[\rm arcsec]$ & $[\kpc]$ & $[\kms]$ & $[{\kms}]$\\
\hline
    0.000 & 0.00 &   3.5 &  5.3 \\
    0.593 & 1.45 & 114.1 &  5.8 \\
    1.185 & 2.89 & 153.8 &  2.9 \\
    1.778 & 4.33 & 212.7 &  2.6 \\
    2.370 & 5.78 & 243.8 &  2.6 \\
    2.963 & 7.22 & 259.8 &  2.3 \\
    3.555 & 8.67 & 256.8 &  2.0 \\
    4.148 & 10.11 & 254.9 &  7.5 \\
    4.740 & 11.56 & 263.4 &  2.3 \\
    5.333 & 13.00 & 265.9 &  3.5 \\
\hline
\hline 
\label{tab:rc}
\end{tabular}
\end{minipage}
\end{table}


\begin{figure*}
  \centering\includegraphics[width=0.85\linewidth]{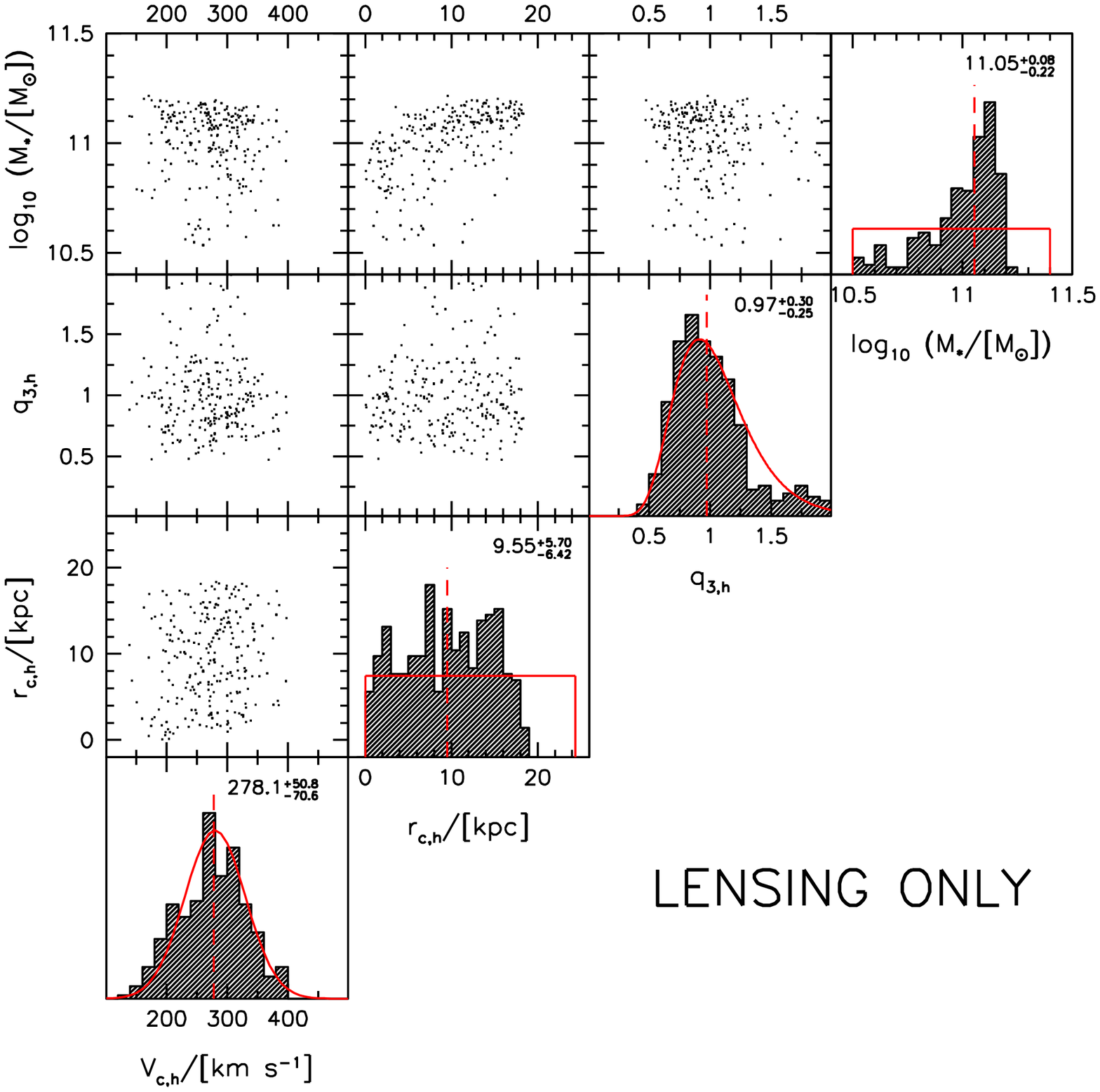}
  \caption{Marginalised two-dimensional posterior PDFs for
    unconstrained mass model parameters (see Table \ref{tab:priors})
    using constraints from strong lensing alone. In the histogram
    panels, the vertical red lines show the median value. The median
    value together with the offsets to the 84th and 16th percentiles
    of the distribution is given in the top right corner. The priors
    are shown with solid red lines.}
\label{fig:lensing}
\end{figure*}

\begin{figure*}
  \centering\includegraphics[width=0.85\linewidth]{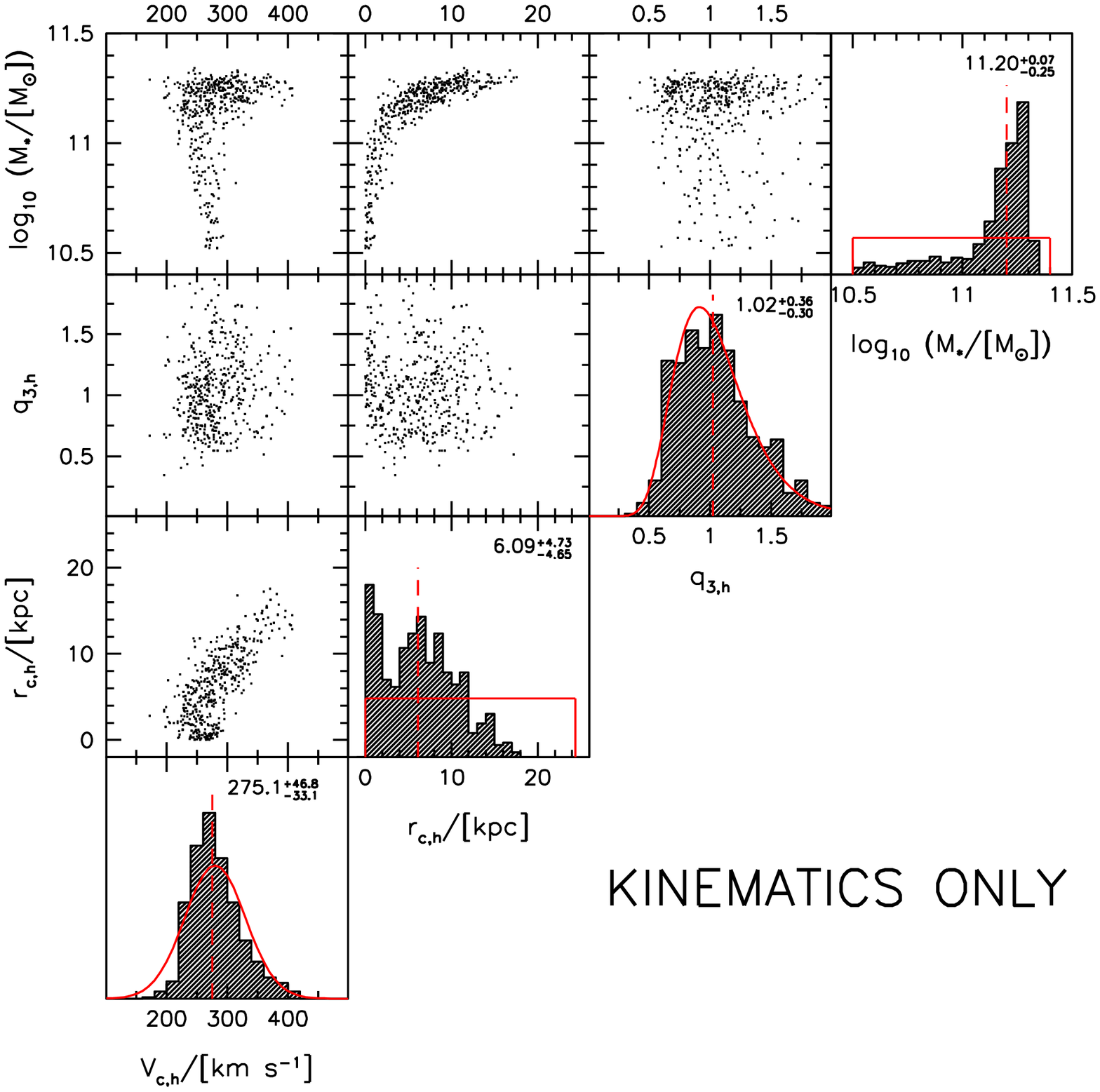}
  \caption{Marginalised two-dimensional posterior PDFs for
    unconstrained mass model parameters (see Table \ref{tab:priors})
    using constraints from kinematics alone. In the histogram panels
    the vertical red lines show the median value. The median value
    together with the offsets to the 84th and 16th percentiles of the
    distribution is given in the top right corner. The priors are
    shown withe solid red lines.}
\label{fig:dynamics}
\end{figure*}

\begin{figure*}
  \centering\includegraphics[width=0.85\linewidth]{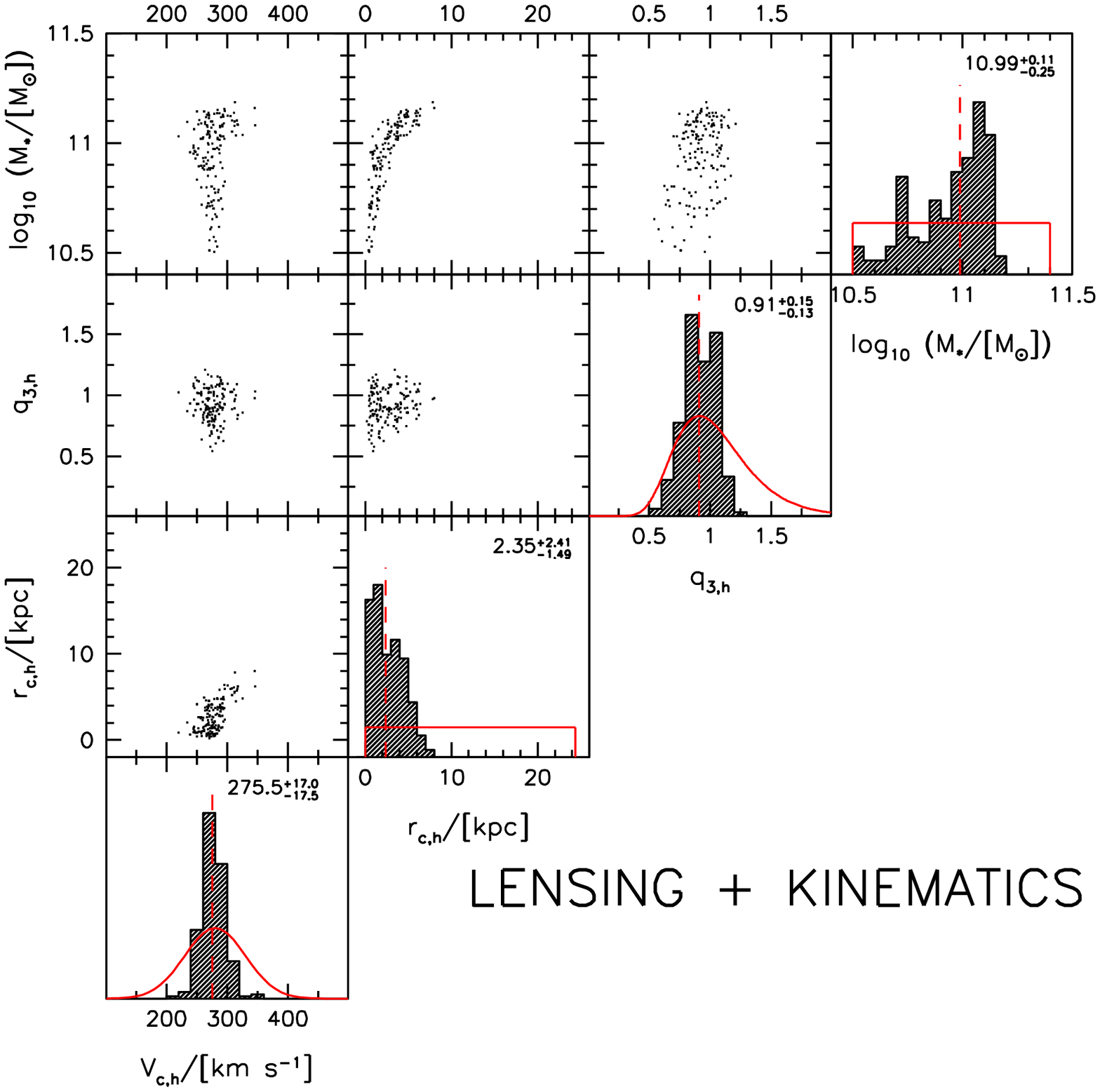}
  \caption{Marginalised two-dimensional posterior PDFs for
    unconstrained mass model parameters (see Table \ref{tab:priors})
    using constraints from both kinematics and strong lensing. In the
    histogram panels the vertical red lines show the median value. The
    median value together with the offsets to the 84th and 16th
    percentiles of the distribution is given in the top right
    corner. The priors are shown with solid red lines.}
\label{fig:joint}
\end{figure*}


\subsection{The central velocity dispersion}
\label{sec:spec:vdisp}

The spectral line fits described in the previous section also yield
some information on the velocity dispersion of the system.  The
central (within 1'') velocity dispersion from the \Mgb--\FeII lines
was found to be $\sigma=180 \pm 4 \kms$, in agreement with the SDSS
value (which is integrated over the 3'' fibre aperture).  \NaD gives a
lower central velocity dispersion, of $\sigma=119\pm 6\kms$.
Absorption in \NaD can come from interstellar gas, as well as
stars. Since \lens has a dusty gas disk, it is thus likely that the
\NaD line is not reliably tracing the stellar velocity dispersion.
For the emission lines the central (within the inner $0.25''$)
velocity dispersion of the \NII line is $\sim 178\pm 14\kms$, similar
to that of the stars. However, the velocity dispersion of the \NII
line declines faster with radius than that of the stars (middle panel
of \Fref{fig:rc}).  This is an indication that the peak in velocity
dispersion in \NII is due to beam-smearing.  For the \Ha line the
central velocity dispersion is considerably lower than that of \NII,
which we ascribe to the presence of absorption, which we have not
corrected for. In the outer part of the disk, the observed velocity
dispersion of the emission lines is close to that of the instrumental
resolution of $\sim 32\kms$ (dotted horizontal line in middle panel of
\Fref{fig:rc}), indicating that the intrinsic velocity dispersion of
the line-emitting gas disk is too low to be resolved.

How could we model the velocity dispersion data?  Our simple dynamical
model does not easily allow for this, but we can make use of the
dispersion information as a cross-check in the following simplistic
way.  The results of Padmanabhan \etal (2004) and Wolf \etal (2010)
show that, for a spherical system, the circular velocity at the
half-light radius $V_{\rm circ}(\Reff)\simeq 1.7 \sigma_{\rm los}$,
where $\sigma_{\rm los}$ is the integrated line of sight velocity
dispersion of the system. For the case of \lens, the {\it bulge} half
light radius is $\Reffb \simeq 0.26"$, and thus the integrated
velocity dispersion within the inner $1"$ gives the integrated
velocity dispersion of the bulge.

Applying the $V_{\rm circ}$ formula above and adopting an uncertainty
of 10\% results in an estimate of the circular velocity at $\Reff$ of
$V_{\rm circ}(R_{50,\rm b})=306\pm31 \kms$. In our current analysis we do
not make use of this constraint. Rather, we use this as a consistency
check to our models which are constrained by strong lensing and gas
rotation curve.


\section{Results}
\label{sec:results}

We now infer the parameters of our 3-component mass model using
constraints from the strong lensing and kinematics data presented in
the previous two sections.  In order to sample the posterior
distribution for the parameters we use the Diffusive Nested Sampling
code from Brewer \etal (2009).  Diffusive Nested Sampling is a
powerful and efficient alternative to standard MCMC sampling.

\begin{figure*}
\centerline{
\psfig{figure=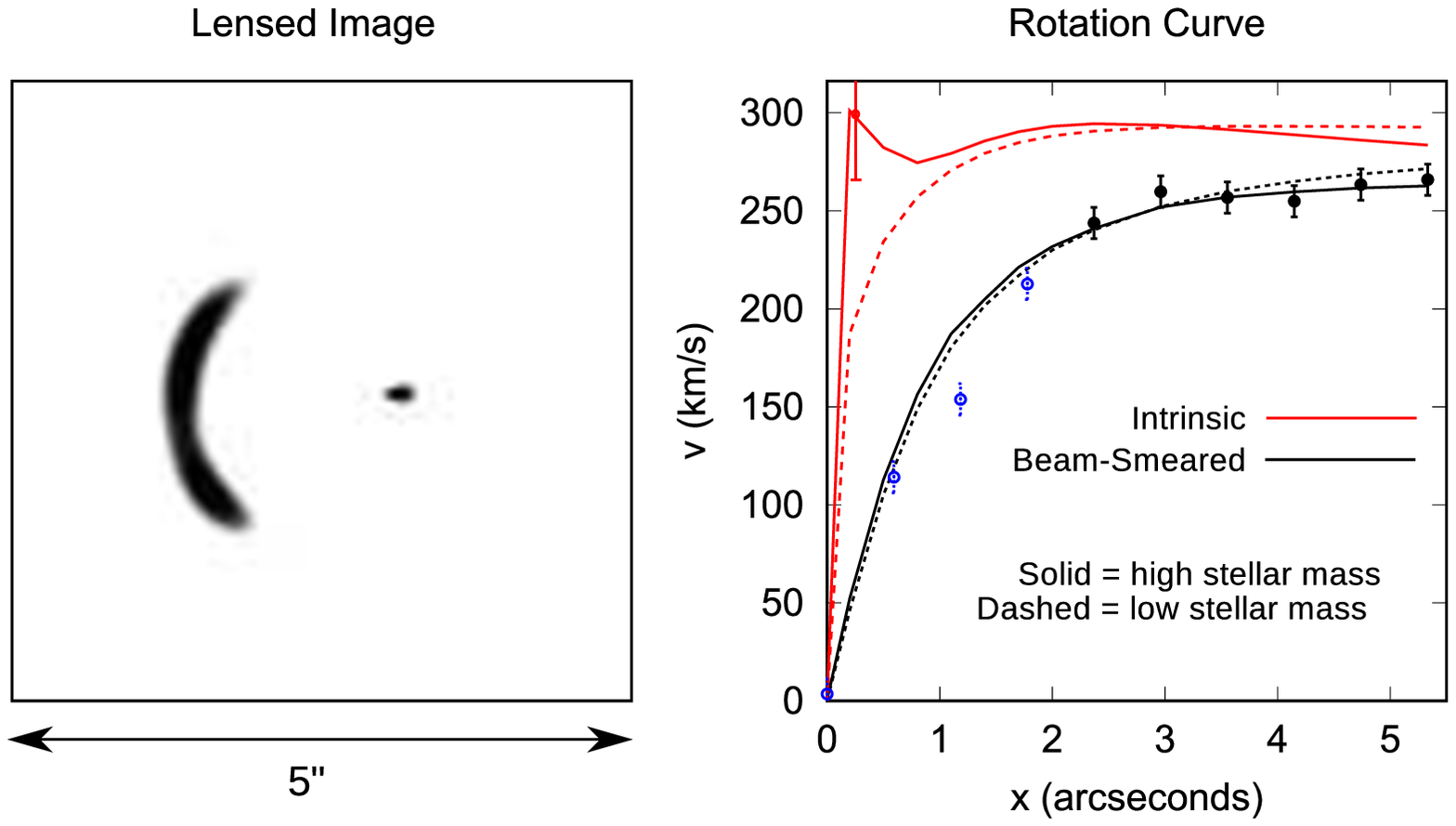,width=0.98\textwidth}
}
\caption{Example mass model that fits the lensing (left panel) and
  kinematics (right panel) data. In the right panel two models are
  shown: median stellar mass (solid lines); low stellar mass (dashed
  lines). The red lines show the intrinsic model circular velocity,
  while the black lines show the model circular velocity after
  beam-smearing, finite slit width, and inclination effects are taken
  into account. Only the black points beyond 2 arcsec are included in
  the fit. The red point at small radii is the constraint from the
  stellar velocity dispersion, and disfavors the low stellar mass
  solution.}
\label{fig:2141mm}
\end{figure*}


\subsection{Inferred Model Parameters}

We consider inferences from three data sets:
\begin{enumerate}
\item strong lensing only
\item kinematics only
\item strong lensing plus kinematics
\end{enumerate}  

In Figures~\ref{fig:lensing}--\ref{fig:joint} we plot, for each
data set, all possible one-dimensional and two-dimensional marginalised
posterior PDFs for the four main mass model parameters. These
parameters are the total (disk+bulge) stellar mass $\Mstar$, and the
dark matter halo asymptotic circular velocity, $\Vhalo$, core radius
$\rhalo$, and 3 dimensional flattening, $\qhalo$.  The median, 16th
and 84th percentiles of the marginalized posterior PDFs for these
parameters individually are given in \Tref{tab:mm}.

The constraints on stellar mass, dark halo density, and dark halo
shape are discussed in more detail below. We first point out some of
the main features of these figures.

With strong lensing alone (\Fref{fig:lensing}), the halo parameters
are poorly constrained: The PDF for the core radius is almost uniform,
while the PDFs for the halo velocity and halo axis ratio follow the
priors. This is expected owing to the limited range in projected
radius probed by the lensing constraints. There is, however, a good
constraint on the stellar mass:
$\log_{10}(\Mstar/\Msun)=11.05^{+0.08}_{-0.22}$.  This is a result of
the axis ratio of the projected mass being quite low (\S
\ref{sec:lensq}).

With kinematics alone (\Fref{fig:dynamics}), the halo core radius is
slightly better constrained, and the stellar mass is less well
constrained. There is a strong degeneracy between the halo core radius
and the stellar mass, with higher stellar masses requiring higher core
radii --- this is the classic disk-halo degeneracy. Related to this
there is a degeneracy between halo velocity and core radius, with
higher halo velocities requiring larger core radii.

\begin{table*}
 \centering
 \begin{minipage}{0.64\linewidth}
   \caption{Summary of fitted parameters: stellar mass ($\Mstar$);
     halo asymptotic circular velocity ($v_{\rm c,h}$); halo core
     radius ($r_{\rm c,h}$); and 3D halo axis ratio ($q_{3,\rm h}$).}
  \begin{tabular}{lccccccc}
   \hline\hline
         & $\log_{10}(\Mstar/\Msun)$ & $V_{\rm c,h}/[\kms]$ & $r_{\rm c,h}/[\kpc]$ & $q_{\rm 3,h}$ \\
\hline
Lensing  & $11.05^{+0.08}_{-0.22}$   & $278^{+51}_{-71}$  & $9.6^{+5.7}_{-6.4}$ & $0.97^{+0.30}_{-0.25}$ \\\\
Kinematics & $11.20^{+0.07}_{-0.25}$   & $275^{+47}_{-33}$    & $6.1^{+4.7}_{-4.7}$  & $1.02^{+0.36}_{-0.30}$ \\\\
Lensing + Kinematics  & $10.99^{+0.11}_{-0.25}$   & $276^{+17}_{-18}$    & $2.35^{+2.4}_{-1.5}$  & $0.91^{+0.15}_{-0.13}$ \\
  \hline
  \hline
\label{tab:mm}
\end{tabular}
\end{minipage}
\end{table*}

Adding the strong lensing constraints to the kinematics constraints
breaks some of the degeneracies. Specifically, it removes the highest
stellar mass solutions from the kinematics only analysis. All
posteriors are considerably tighter than the priors, illustrating the
power of the combined analysis: for example, circular velocity is now
known to 6\% precision, and core radius is well constrained to be
smaller than 5 kpc. However, there is still a degeneracy between halo
core radius and stellar mass.  There is also a residual degeneracy
between stellar mass and halo shape --- with low stellar mass
solutions favoring oblate dark matter haloes. This degeneracy is
expected as the total mass needs to be flattened to reproduce the
strong lensing (\Sref{sec:lensq}). The flattening can be achieved with
either a significant stellar disk component and a spherical halo, or a
less massive disk and a more flattened halo.

In \Fref{fig:2141mm} we show the rotation curves and strong lensing
image predicted by two example mass models drawn from the posterior
PDF given both lensing and kinematics data. These models both predict
four images of the lensed source, including a faint counter-image that
is consistent with the noise in the centre of the subtracted image.
Both models' predicted rotation curves fit the outer part of the
observed rotation curve very well; they also match well the 
central part of the observed rotation curve, which was not used in the
fit. The two models have either the posterior median stellar mass, or
much lower stellar mass; they can only be distinguished in the plot of
the intrinsic, pre beam-smeared rotation curves, where the high
stellar mass model has a significantly higher rotation velocity at
radii less than one arcsec. This region could be probed with higher
spatial resolution spectroscopy, or by making use of the velocity
dispersion information. Indeed, our cross-check point from
\Sref{sec:spec:vdisp} would favour the high stellar mass model.

\begin{figure*}
\centerline{
\psfig{figure=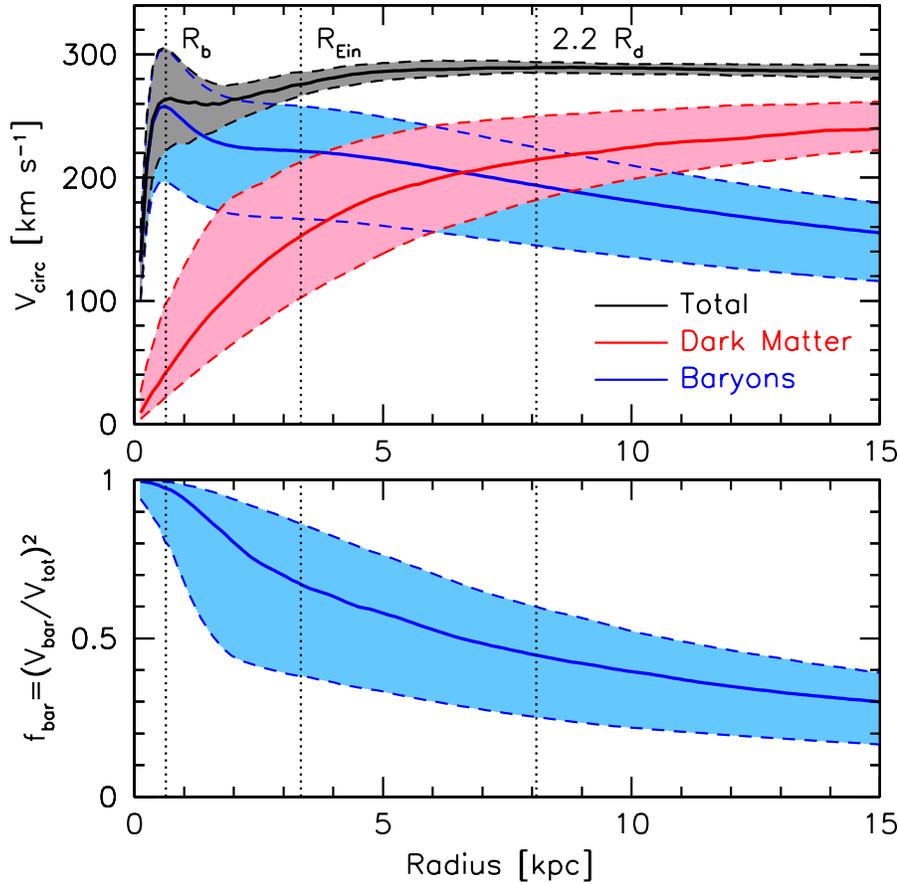,width=0.70\textwidth}
}
\caption{Circular velocity profiles (upper panel) and spherical baryon
  fractions (lower panel) from our joint lensing plus kinematics
  analysis. The solid lines show the median, while the shaded regions
  enclose 68\% of the posterior PDF. The total circular velocity
  (black line and grey shaded region) is well constrained outside of
  the Einstein radius, $R_{\rm Ein}$, and up to the last rotation curve
  point at 13 kpc. The contributions of the baryons (red lines and
  shaded regions) and the dark matter (blue lines and shaded regions)
  are more uncertain. However, at the bulge half-light radius,
  $R_{b}$, the galaxy is baryon dominated (and thus is ``maximal''),
  while at 2.2 disk scale lengths the baryons fraction is roughly 50\%
  (and thus is ``sub-maximal''). \label{fig:mass_within}}
\end{figure*}

In Figure~\ref{fig:mass_within} we show the inferred circular velocity
profile, decomposed into baryonic and dark matter components. These
estimates are based on the posterior samples using the joint lensing
plus kinematics analysis. The solid lines show the median model from the
posterior PDF, while the shaded regions enclose 68\% of the models. In
the radial region where we have observational constraints (i.e., from
the Einstein radius to the last rotation curve point) the total
circular velocity is well constrained. For example, at 2.2 disk scale
lengths (8.1 kpc), the total circular velocity is
$V_{2.2}=289\pm4\kms$. The circular velocity profiles of the baryons
and dark matter are not as tightly constrained. Nevertheless, we can
still infer interesting constraints on the dark matter fraction as a
function of radius, and thus determine whether or not \lens has a
maximum disk.  A working definition of a maximum disk is when $V_{\rm
  disk}(2.2 R_{\rm d})/V_{\rm tot}(2.2 R_{\rm d})=0.85\pm0.10$
(Sackett 1997). Here $V_{\rm disk}(2.2 R_{\rm d})$ is the circular
velocity of the disk at 2.2 disk scale lengths, and $V_{\rm tot}(2.2
R_{\rm d}) \equiv V_{2.2}$ is the total circular velocity at 2.2 disk
scale lengths.

A galaxy may have a sub-maximal disk, but still have a maximal
baryonic component due to the bulge. Thus we consider the contribution
of the baryons (i.e., bulge plus disk) to $V_{2.2}$ to be of more
relevance than just the contribution of the disk to $V_{2.2}$. We find
that $V_{\rm bar}(2.2R_{\rm d})/V_{2.2}=0.67^{+0.10}_{-0.17}$, 
which suggests that \lens is sub-maximal at 2.2 disk scale
lengths. However, the baryon contribution to the total circular
velocity increases towards smaller radii (lower panel of
Fig.~\ref{fig:mass_within}) such that $V_{\rm bar}(R_{\rm b})/V_{\rm
  tot}=0.99^{+0.01}_{-0.09}$, and thus \lens is maximal at the bulge
half-light radius. Converting circular velocities into spherical
masses, results in a dark matter fraction of $f_{\rm
  DM}=0.55^{+0.20}_{-0.15}$ within 2.2 disk scale lengths, and $f_{\rm
  DM}=0.02^{+0.17}_{-0.02}$ within the bulge half-light radius.

\subsection{Constraints on the stellar IMF}
\label{sec:imf}
\begin{figure}
\centerline{
\psfig{figure=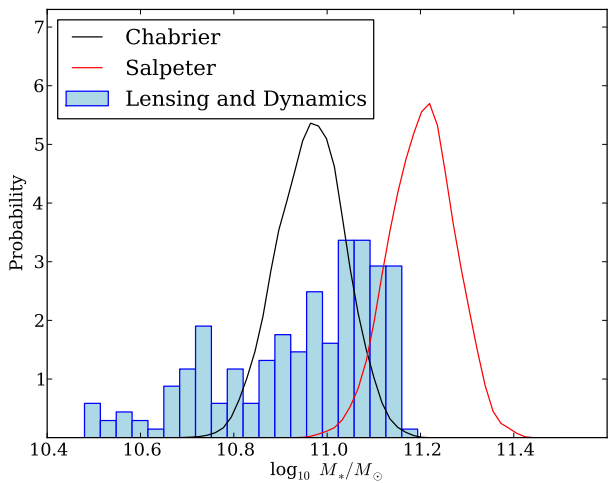,width=0.45\textwidth}
}
\caption{Inference on stellar mass from lensing and kinematics
  (histogram) compared with SPS models (solid lines) assuming a
  Chabrier IMF (black) and Salpeter IMF (red). The Bayes factor in
  favor of a Chabrier IMF, compared to a Salpeter IMF is
  2.5.}
\label{fig:mstarhist}
\end{figure}

\Fref{fig:mstarhist} shows the posterior PDFs from our joint lensing
and kinematics analysis together with those from SPS models for both
Chabrier and Salpeter IMFs. From our lensing and kinematics analysis
the stellar mass of the galaxy is found to be
$\log_{10}\Mstar=10.99^{+0.11}_{-0.25}$.  This is in excellent
agreement with the stellar mass derived from SED fitting assuming a
Chabrier IMF, which is $\log_{10}\Mstar=10.97\pm{0.07}$. A Salpeter
IMF results in stellar masses 0.24 dex higher. Our analysis thus
mildly favors a Chabrier IMF over a Salpeter IMF.  We can quantify
this agreement by integrating the likelihood over a prior for the
stellar mass defined by either the Chabrier or the Salpeter SPS model
PDFs.  The ratio of these integrals is the Bayes factor, or evidence,
in favour of a Chabrier IMF; we find its value to be 2.7, which is to
say that the data are 2.7 times more likely to have come from the
Chabrier model than from the Salpeter one. If these are the only two
models possible, then there is a 73\% chance that the Chabrier model
is the true one. This corresponds to weak evidence in favor of
Chabrier vs Salpeter.

\begin{figure}
\centerline{
\psfig{figure=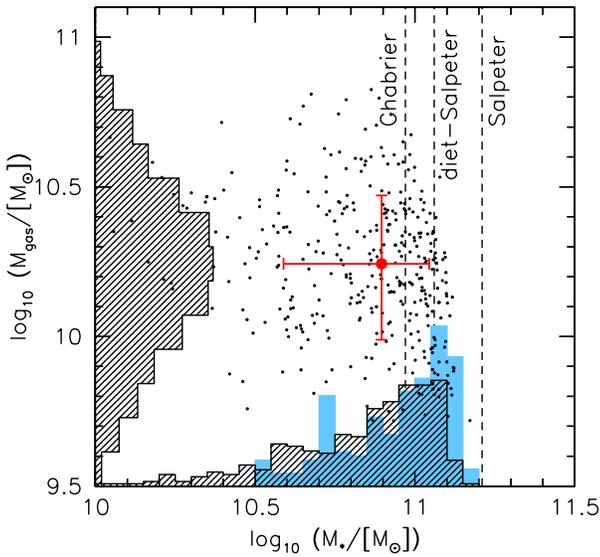,width=0.45\textwidth}
}
\caption{Effect of gas mass on the inferred stellar mass from lensing
  and kinematics. For each model galaxy in the posterior PDF we draw a
  gas mass from a log-normal distribution with mean and standard
  deviation typical for massive spiral galaxies. The resulting PDFs
  for the stellar and gas mass are shown as black hatched
  histograms. For comparison, the blue shaded histogram shows the
  posterior PDF on the stellar mass assuming no gas mass. Thus
  accounting for cold gas mass reduces the stellar mass derived from
  lensing and kinematics by $\simeq 0.1$ dex.} 
\label{fig:mstarhist2}
\end{figure}

In our current analysis we ignore the possibility of cold gas. For
massive spiral galaxies the cold gas fraction is $\simeq 20\pm10\%$
(assuming a Chabrier IMF), split roughly equally between atomic and
molecular gas (e.g., Dutton \& van den Bosch 2009).  If the cold gas
is distributed like the stars, then the lensing+kinematics stellar
mass is actually a baryonic mass, greater than or equal to the actual
stellar mass. If the cold gas is more extended than the stars, as is
often the case, then we will still be over-estimating the stellar
mass, but by a smaller amount.
To estimate an upper limit to the impact of cold gas on our derived
stellar masses we assume that the gas mass for \lens is distributed
like the stars.  For each model in the posterior PDF we draw a gas
mass from a log-normal distribution centered on $M_{\rm gas}=1.8\times
10^{10}\Msun$, with a standard deviation of 0.3 dex. We then subtract
off the gas mass from the gas free stellar mass to derive the ``true''
stellar mass. The results of this exercise are shown in
Fig.~\ref{fig:mstarhist2}. The resulting median and 68\% confidence
interval on the stellar mass is
$\log_{10}(\Mstar/\Msun)=10.89^{+0.15}_{-0.33}$, i.e., 0.1 dex lower
than when ignoring the cold gas.  The Bayes factor in favor of a
Chabrier IMF over a Salpeter IMF has increased from 2.7 to 11.9, which
corresponds to strong evidence.  Thus by ignoring the cold gas we
could be over estimating the stellar mass by $\simeq 0.1\pm0.05$ dex,
which strengthens the case for an IMF lighter than Salpeter.

How does this result compare to previous work?  Using maximal disk
fits to spiral galaxy rotation curves in the Ursa Major cluster, Bell
\& de Jong (2001) placed an upper limit on the stellar mass-to-light
ratio normalisation, favoring IMFs with stellar masses 0.15 dex lower
than Salpeter, the so-called diet-Salpeter IMF. We note that a
Salpeter IMF is also disfavored for fast rotating elliptical galaxies
(Cappellari \etal 2006; Treu et al. 2010; Auger et al. 2010; Barnab\'e
et al.\ 2010), but is favored for massive elliptical galaxies (Treu
\etal 2010; Auger \etal 2010; van Dokkum \& Conroy 2010). Thus
comparing our result with those for massive ellipticals, supports the
idea that the IMF is not universal, but dependent on galaxy mass
and/or Hubble type.

By shifting our Salpeter stellar mass PDF by $-0.15$~dex, we find that
for \lens a diet Salpeter IMF corresponds to a stellar mass of
$\log_{10}(\Mstar/\Msun)=11.06\pm{0.07}$. This IMF is favored over a
Salpeter IMF, by Bayes factors of 3.5 (assuming no cold gas), and 9.9
(assuming a gas mass of $\log_{10}(M_{\rm gas}/\Msun)=10.26\pm0.30$). 
There is little evidence distinguishing
between diet Salpeter and Chabrier IMFs.


\subsection{Constraints on the dark halo density profile}
\label{sec:halodensity}
N-body simulations have shown that in $\Lambda$CDM cosmologies dark
matter haloes are expected (in the absence of baryonic effects) to
have very specific structure. The mass density profiles can be well
approximated by the so called NFW profile (Navarro \etal 1997). This
has a density profile that varies from $\rho(r) \propto r^{-1}$ at
small radii, to $\rho \propto r^{-3}$ at large radii.  The radius
where the logarithmic slope of the density profile is ${\rm
d}\ln\rho/{\rm d}\ln r=-2$ is known as the scale radius, $r_{\rm s}$.
The halo scale radii are tightly correlated with the virial masses of
dark matter haloes, $M_{\rm vir}$. This correlation is usually
expressed in terms of the halo concentration, $c=r_{\rm vir}/r_{\rm s}$,
where $r_{\rm vir}$ is the virial radius. Halo concentrations are only
weakly dependent on halo mass, with a relation of the form $c\propto
M_{\rm vir}^{-0.1}$ (Macci\'o \etal 2007).  The scatter in halo
concentration, at fixed halo mass, for relaxed haloes is small $\simeq
0.11$ dex (Jing 2000; Wechsler \etal 2002; Macci\'o \etal
2007).

\begin{figure}
\centerline{
\psfig{figure=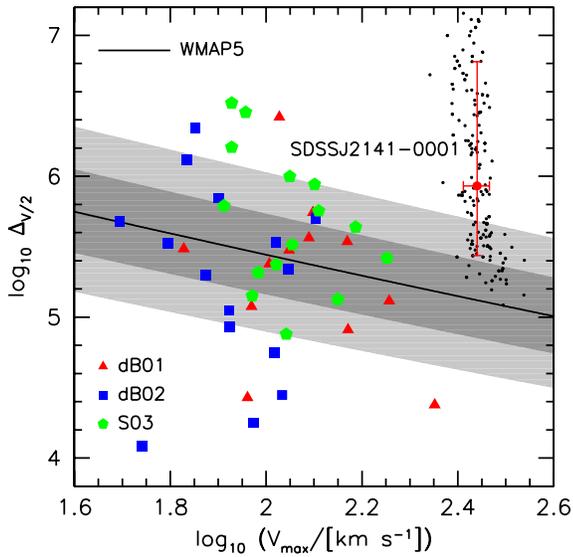,width=0.45\textwidth}
}
\caption{Central density of the dark matter halo, $\Delta_{V/2}$ vs
  maximum halo circular velocity, $V_{\rm max}$. The solid black line
  shows the prediction for pristine dark matter haloes in the
  concordance $\Lambda$CDM cosmology (WMAP5) from Macci\`o \etal
  (2008). The shaded regions show the 1 and $2\sigma$ intrinsic
  scatter.  The colored symbols show measurements from dwarf and low
  surface brightness galaxies, after subtraction of the baryons (de
  Blok \etal 2001, dB01; de Blok \& Bosma 2002, dB02; Swaters \etal
  2003, S03). The small black dots show samples from the posterior PDF
  for \lens. The large red dot and error bars show the median and 68\%
  ranges of this PDF, respectively.}
\label{fig:deltaV2}
\end{figure}

Observationally, measuring halo concentrations is a challenge because
halo virial masses are poorly constrained for individual galaxies, due
to the lack or sparsity of dynamical tracers at large radii.  A more
observationally accessible measure of dark halo structure is the
parameter $\Delta_{V/2}$, which depends on the maximum halo circular
velocity, $V_{\rm max}$, and the radius where circular
velocity of the halo is half of the maximum, $r_{V/2}$ (Alam, Bullock
\& Weinberg 2002):
\begin{equation}
  \Delta_{V/2}=5\times 10^5 %
  \left(\frac{V_{\rm max}/[100 \kms]} {r_{V/2}/[h^{-1}\kpc]}\right)^2.
\end{equation}
For NFW haloes there is a one-to-one mapping between
$\Delta_{V/2}-V_{\rm max}$ and $c-M_{\rm vir}$, and thus one can
compare the observed $\Delta_{V/2}$ with predictions for $\Lambda$CDM
haloes.  \Fref{fig:deltaV2} shows the predictions for
$\Delta_{V/2}-V_{\rm max}$ in a WMAP 5th year cosmology (Dunkley \etal
2009) from Macci\`o, Dutton, \& van den Bosch (2008). The shaded
regions show the 1 and 2$\sigma$ intrinsic scatter.  The large symbols
show measurements from dwarf and low surface brightness galaxies after
subtracting of the baryons (de Blok \etal 2001; de Blok \& Bosma 2002;
Swaters \etal 2003). These are in excellent agreement with the
predictions from $\Lambda$CDM.

\begin{figure}
\centerline{
\psfig{figure=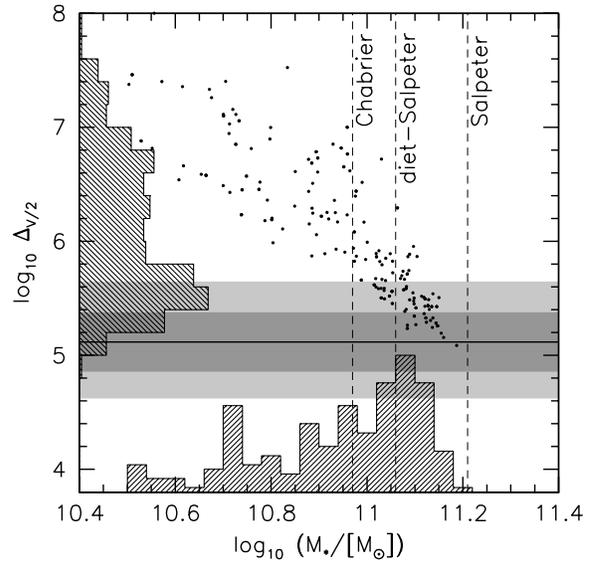,width=0.45\textwidth}
}
\caption{Central density of the dark matter halo, $\Delta_{V/2}$ vs
  stellar mass, $\Mstar$. The solid black line shows the prediction
  for pristine dark matter haloes in the concordance $\Lambda$CDM
  cosmology (WMAP5). The shaded regions show the 1 and $2\sigma$
  intrinsic scatter.  The black dots show samples from the posterior
  PDF for \lens. The maximum likelihood stellar masses from SPS models
  assuming Chabrier, diet-Salpeter, and Salpeter IMFs are shown with
  vertical dashed lines.}
\label{fig:deltaV2_mstar}
\end{figure}

In our mass model of \lens, the dark matter halo has a softened
isothermal density profile, which has $V_{\rm max}=V_{\rm c,h}$, and
$r_{V/2}=1.1263 \,r_{\rm c,h}$.  We can compute $\Delta_{V/2}$ for
this model and compare it with the NFW profile halos of the
simulations. For our model the median and uncertainty (corresponding
to 16th and 84th percentiles) is $\log_{10}\Delta_{V/2}
=5.9^{+0.9}_{-0.5}$. 
The median is $2.7\sigma$ higher (in terms of intrinsic scatter) than
that predicted for pristine $\Lambda$CDM haloes, although the full
posterior PDF overlaps the $\Lambda$CDM predictions, as shown in
\Fref{fig:deltaV2}. The 16th percentile of the PDF for $\Delta_{V/2}$
only corresponds to a $1.1\sigma$ deviation from the $\Lambda$CDM
distribution. Thus there is a suggestion that the \lens halo is higher
density than expected.  We note that, as shown in
Fig.~\ref{fig:deltaV2_mstar}, the central density is highly correlated
with the stellar mass, with lower $\Delta_{V/2}$ for higher $\Mstar$.

There are two interpretations of the higher than expected halo
density. 1) The halo has undergone contraction in response to galaxy
formation (e.g., Blumenthal \etal 1986). 2) The halo of \lens has a
higher density than typical haloes of the same mass.  In order to
distinguish between these two scenarios it is necessary to understand
the selection function of the SWELLS lens galaxies. We cannot do this
with one galaxy, and therefore we will leave this for future
work. Nevertheless, we can gain some insight by investigating where
\lens falls on disk galaxy scaling relations.

We consider the relations between stellar mass ($\Mstar$), rotation
velocity at 2.2 disk scale lengths ($V_{2.2}$), and disk scale length
($R_{\rm d}$).  For \lens the values of these parameters are
$V_{2.2}\simeq 289 \kms$, $\log_{10}(M_{*,\rm Chab}/\Msun) \simeq
11.0$, and $R_{\rm d} \simeq 3.7 \kpc$.  For a stellar mass of
$10^{11}\Msun$ we expect $V_{2.2}=229\pm 25\kms$ (Dutton \etal 2010b).
Alternatively, for a rotation velocity of $V_{2.2}=289 \kms$ we expect
a stellar mass of $\log_{10}(\Mstar/\Msun)=11.39^{+0.18}_{-0.18}$,
assuming a Chabrier IMF.  For a rotation velocity of $V_{2.2}=289
\kms$ we expect $R_{\rm d}=6.0^{+2.7}_{-1.9} \kpc$ (Courteau \etal
2007; Dutton \etal 2007). Thus \lens is offset to low stellar mass (by
$\simeq 2\sigma$ and small size (by $\simeq 1.3\sigma$) at fixed
rotation velocity, which means that it has a higher baryonic and total
mass density than typical massive spiral galaxies.

The high central density of \lens may be related to how it was
selected.  If the central densities of spiral galaxies are close to
the critical value for strong lensing, then the galaxies that are
observed to be strong lenses could be a biased sub-set of the
population. Thus the conclusions we draw for \lens might not
necessarily be applicable to spiral galaxies in general. This
selection effect is likely to affect the interpretation of some
parameters more than others. For example, in order to interpret the
central densities of dark matter haloes in terms of the halo response
to galaxy formation it is necessary for the lenses to reside in an
unbiased subset of haloes. However, if we assume that the IMF is
universal, at least for massive spiral galaxies, then the constraints
we place on the IMF using spiral galaxy strong lenses will be
independent of any selection bias.  Larger samples of spiral galaxy
lenses are obviously needed to fully characterise the selection biases
in spiral galaxy lenses.


\subsection{Constraints on the dark halo shape}
\label{sec:haloshape}
N-body simulations have shown that in $\Lambda$CDM cosmologies dark
matter haloes are triaxial, with a preference towards prolate shapes
(Jing \& Suto 2002; Bailin \& Steinmetz 2005; Kasun \& Evrard 2005;
Allgood \etal 2006; Bett \etal 2007; Macci\`o \etal 2007, 2008). For a
halo mass of $M_{\rm vir}=10^{12}\Msun$, which is typical for massive
spiral galaxies, the minor to major axis ratio ratio $c/a\simeq
0.63\pm 0.1$, and the intermediate to major axis ratio $b/a\simeq
0.80\pm 0.1$ (Macci\`o \etal 2008). $\Lambda$CDM haloes are also found
to be more prolate at smaller radii (Allgood \etal 2006; Macci\`o
\etal 2008; Abadi \etal 2010) .

The assembly of a central galaxy is expected to modify the three
dimensional shape of the dark matter halo (Katz \& Gunn 1991; Dubinski
1994).  Using cosmological simulations Abadi \etal (2010) found that
as a result of the assembly of the central galaxy the haloes become
nearly oblate with $b/a \simeq 0.95$ and $c/a\simeq 0.85$, and the
axial ratios become approximately independent of radius. Similar
results have been obtained from other cosmological simulations (e.g.,
Tissera \etal 2010). It should be noted that these simulations
over-predict the baryon to dark halo mass ratios, and thus the effect
of galaxy assembly on the halo shapes could be over-estimated.

In our mass models we assume the halo is axisymmetric, with 3D axis
ratio $\qhalo$. A spherical halo has $\qhalo=1$, an oblate halo has
$\qhalo<1$, and a prolate halo has $\qhalo > 1$.  We adopt a
log-normal prior on the halo axis ratio, centered on $\qhalo=1$. For
our fits to lensing only or kinematics only the posterior PDF for
$\qhalo$ is identical to the prior, but for the joint analysis
slightly oblate haloes are preferred: $\qhalo=0.91^{+0.15}_{-0.13}$,
and prolate haloes with $\qhalo > 1.2$ are strongly disfavored. Thus
our results for \lens support the notion that galaxy assembly
sphericalizes the dark matter halo, and perhaps even flattens it
towards the disk.


\section{Conclusions}
\label{sec:concl}

We have presented an analysis of the strong gravitational lens
\lens, discovered as part of the SLACS survey, using data
from {\it HST} and the Keck telescopes. The lens galaxy is a high
inclination disk dominated galaxy with $K'$-band bulge fraction of
0.2, showing stellar rotation in multiple spectral lines.
A singular isothermal ellipsoid lens model provides a circular
velocity of $V_{\rm c}=254^{+15}_{-18}\kms$ and an axis ratio of
$q=0.42^{+0.17}_{-0.12}$.

We perform a joint fit to the multi-filter surface brightness, lensing
and kinematics  data using a self-consistent 3-component mass model,
and from it draw the following conclusions:

\begin{itemize}

\item The lensing and kinematics constraints yield a stellar mass of
  $\log_{10}(\Mstar/\Msun) = 10.99^{+0.11}_{-0.25}$ (68\% confidence
  interval), independent of the IMF.

\item This value is in excellent agreement with the stellar mass
  derived from the SED using SPS models and assuming a Chabrier (2003)
  IMF: $\log_{10}(\Mstar/\Msun) = 10.97^{+0.07}_{-0.07}$. A Salpeter
  (1955) IMF results in stellar masses 0.24 dex higher: our analysis
  marginally favors a Chabrier IMF over a Salpeter IMF, by a Bayes
  factor of 2.7.

\item Accounting for the expected gas mass reduces the lensing and
  kinematics stellar mass by $0.10\pm0.05$ dex, and increases the
  Bayes factor in favor of a Chabrier IMF to 11.9.

\item At 2.2 disk scale lengths the spherical dark matter fraction is
  $f_{\rm DM}=0.55^{+0.20}_{-0.15}$, suggesting that the baryons are
  sub-maximal.

\item The dark matter halo has a maximum circular velocity of
  $\Vhalo=276^{+17}_{-18} \kms$, and a core radius of
  $\rhalo=2.4^{+2.4}_{-1.5} \kpc$. The corresponding central density
  parameter $\log_{10}\Delta_{V/2}=5.9^{+0.9}_{-0.5}$ is higher than
  expected for uncontracted NFW haloes in the concordance $\Lambda$CDM
  cosmology, which have $\log_{10}\Delta_{V/2}=5.2$ and an intrinsic
  scatter of 0.3.

\item This high density could either be evidence for halo contraction
  in response to galaxy formation (e.g., Blumenthal \etal 1986), or
  the result of a selection bias towards high concentration haloes. A
  larger sample with well-characterised selection function is required
  to make further progress.

\item The dark matter halo is oblate, $\qhalo = 0.91^{+0.15}_{-0.13}$,
  with a probability of 69\%.  This finding provides support for the
  notion that galaxy assembly turns strongly prolate triaxial dark
  matter haloes into roughly oblate axisymmetric haloes (e.g., Abadi
  \etal 2010).

\end{itemize}


\section*{Acknowledgements}
AAD acknowledges financial support from a CITA National Fellowship,
from the National Science Foundation Science and Technology Center
CfAO, managed by UC Santa Cruz under cooperative agreement
No. AST-9876783. AAD and DCK were partially supported by NSF grant AST
08-08133, and by HST grants AR-10664.01-A, HST AR-10965.02-A, and HST
GO-11206.02-A.
PJM was given support by the TABASGO and Kavli foundations in the form
of two research fellowships.
TT acknowledges support from the NSF through CAREER award NSF-0642621,
and from the Sloan Foundation through a Sloan Research Fellowship.
LVEK acknowledges the support by an NWO-VIDI programme subsidy
(programme number 639.042.505).
This research is supported by NASA through Hubble Space Telescope
programs GO-10587 and GO-11978, and in part by the National Science
Foundation under Grant No. PHY99-07949. and is based on observations
made with the NASA/ESA Hubble Space Telescope and obtained at the
Space Telescope Science Institute, which is operated by the
Association of Universities for Research in Astronomy, Inc., under
NASA contract NAS 5-26555, and at the W.M. Keck Observatory, which is
operated as a scientific partnership among the California Institute of
Technology, the University of California and the National Aeronautics
and Space Administration. The Observatory was made possible by the
generous financial support of the W.M. Keck Foundation. The authors
wish to recognize and acknowledge the very significant cultural role
and reverence that the summit of Mauna Kea has always had within the
indigenous Hawaiian community.  We are most fortunate to have the
opportunity to conduct observations from this mountain.
Funding for the SDSS and SDSS-II was provided by the Alfred P. Sloan
Foundation, the Participating Institutions, the National Science
Foundation, the U.S. Department of Energy, the National Aeronautics
and Space Administration, the Japanese Monbukagakusho, the Max Planck
Society, and the Higher Education Funding Council for England. The
SDSS was managed by the Astrophysical Research Consortium for the
Participating Institutions. The SDSS Web Site is http://www.sdss.org/.


\appendix

\section{The Chameleon approximation to a S\'ersic Profile}
\label{app:chameleon}

In this appendix we derive an approximation to a S\'ersic profile as the
difference of two non-singular isothermal ellipsoids (NIE's). 

The S\'ersic profile is specified by three parameters: a normalization,
a radial scale, and a shape parameter commonly known as the S\'ersic
index, $n$. In its simplest form it is given by:
\begin{equation}\label{eq:sersic1}
\Sigma(R) = \Sigma_{0} \exp \left[  - \left( \frac{R}{R_{\rm 0}} \right) ^{1/n} \right]
\end{equation}
where $\Sigma_0$ is the central surface density, and $R_0$ is the
radial scale.  The S\'ersic profile is commonly expressed in terms of
the effective radius, $R_{\rm e}$, which encloses half of the
projected mass, and the effective density, $\Sigma_{\rm e}\equiv
\Sigma(R_{\rm e})$:
\begin{equation}
\label{eq:sersic2}
\Sigma(R) = \Sigma_{\rm e} \exp \left\{ -b_n \left[\left( \frac{R}{R_{\rm e}} \right) ^{1/n} -1 \right] \right\}.
\end{equation}
Where the term $b_n\approx 2 n -0.32$. Here we use the asymptotic
expansion of Ciotti \& Bertin (1999) to $O(n^{-5})$ valid to one part
in $\sim10^{4}$ for $n>0.36$:
\begin{eqnarray}
b_n = 2n - \frac{1}{3} + \frac{4}{405n} +\frac{46}{25515n^2} +\frac{131}{1148175n^3}  \nonumber \\
-\frac{2194697}{30690717750 n^4} + O(n^{-5}).
\end{eqnarray}
\Eref{eq:sersic1} and \Eref{eq:sersic2} are related to each other via
\begin{equation}
R_{\rm e} = (b_n)^n R_0
\end{equation}
and
\begin{equation}
\Sigma_{\rm e} = \exp(-b_n) \Sigma_0.
\end{equation}

The Chameleon profile is the difference between two NIEs with
different core radii, but the same normalization (which insures the
total mass is finite):
\begin{eqnarray}
\Sigma_{\rm chm}(R; \Sigma_1,R_1,R_2)= \nonumber \\
\left( \frac{\Sigma_{1}}{\sqrt{R^2 + R_1^2}} -\frac{\Sigma_1}{\sqrt{R^2 + R_2^2}} \right). 
\end{eqnarray}
For the purpose of comparing to a S\'ersic profile, we use the
following parametrization.
\begin{eqnarray}
\Sigma_{\rm chm}(R; \Sigma_0, R_0,\alpha)= \nonumber \\ 
 \frac{\Sigma_0} {1-\alpha} \left( \frac{R_0}{\sqrt{R^2 + R_0^2}} -\frac{R_0}{\sqrt{R^2 + (R_0/\alpha)^2}} \right). 
\end{eqnarray}
Where $\Sigma_0 = \Sigma_1 (1-\alpha) /R_1$ is the central surface
density of the chameleon profile, $R_0=R_1$, and $\alpha$ is the ratio
between the core radii of the two NIE's: $\alpha=R_1/R_2$, so that $ 0
< \alpha < 1$. The total mass of the Chameleon profile is
\begin{equation}
M_{\rm Chm}=\pi \Sigma_0 R_0^2 /\alpha.
\end{equation}

\begin{figure}
\centering\includegraphics[width=0.95\linewidth]{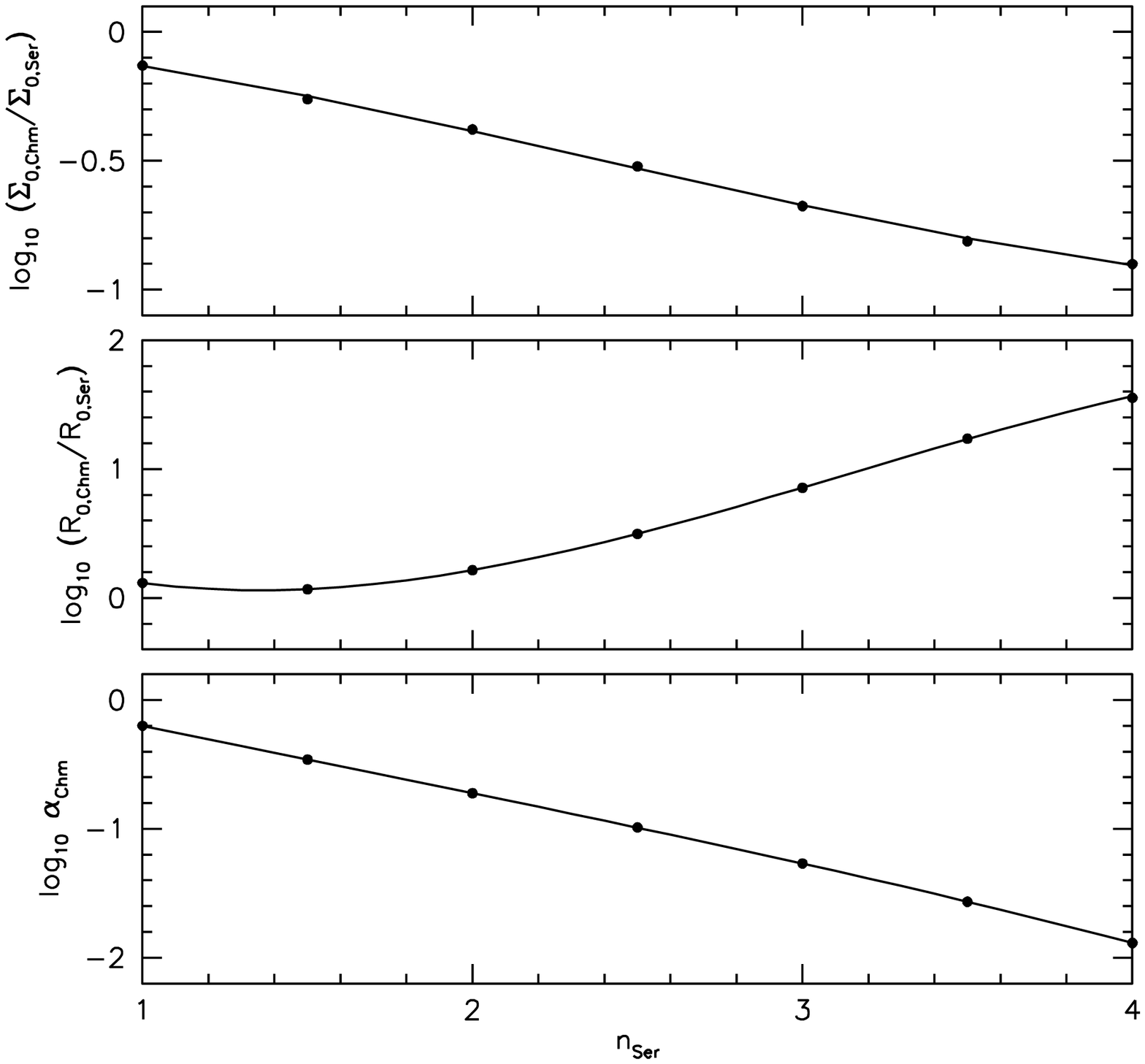}
\caption{Relation between Chameleon and S\'ersic parameters as a
  function of S\'ersic index.}
\label{fig:chmser}
\end{figure}

We wish to find the correspondence between the S\'ersic parameters,
$\Sigma_{0,\rm Ser}$, $R_{0,\rm Ser}$, $n_{\rm Ser}$, and the Chameleon
parameters, $\Sigma_{0,\rm Chm}$, $R_{0,\rm Chm}$, $\alpha_{\rm Chm}$. We do this by
fitting a Chameleon profile to a S\'ersic profile. We are interested
in mass profiles, so we fit $R \,\Sigma(R)$ to give appropriate weight
to the density profile at large radii.  \Fref{fig:chmser} shows
the best fit parameters as a function of S\'ersic index.  We fit these
relations between chameleon and S\'ersic parameters with a cubic
function:
\begin{equation}
\label{eq:power2}
y = y_0 + y_1 (x-x_0)  + y_2 (x-x_0)^2 + y_3 (x-x_0)^3
\end{equation}
The parameters of these fits are given in \Tref{tab:chmserfit}.
Our fitting function is valid for $1\le n_{\rm Ser} \le 4$.

\begin{figure*}
\centering\includegraphics[width=0.99\linewidth]{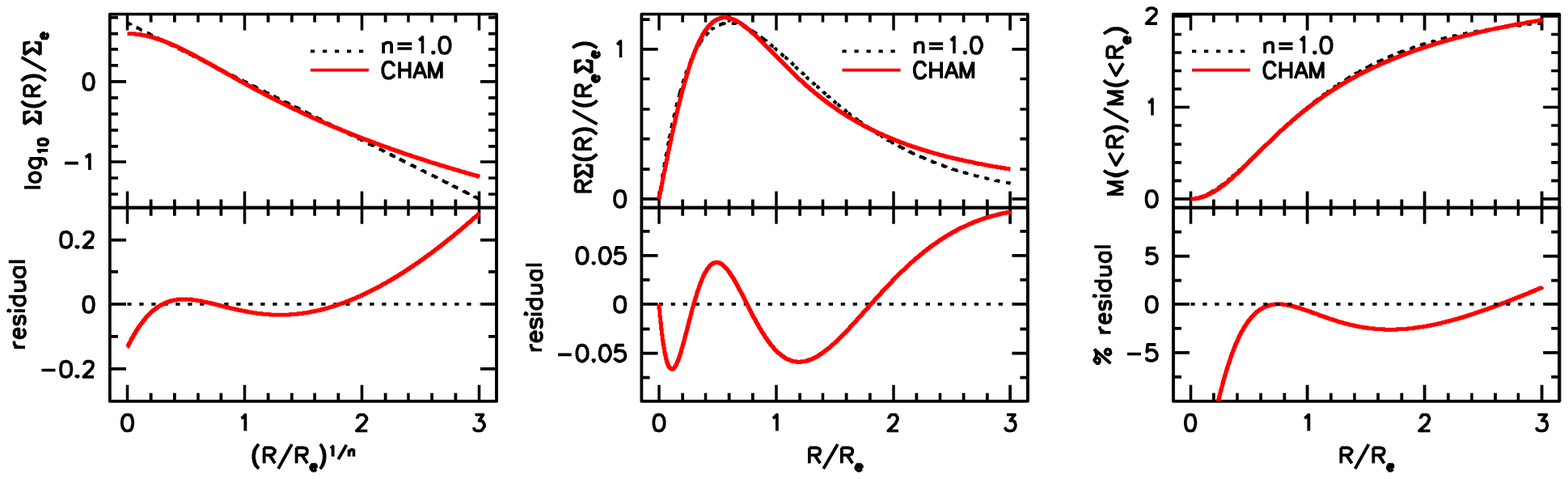}
\centering\includegraphics[width=0.99\linewidth]{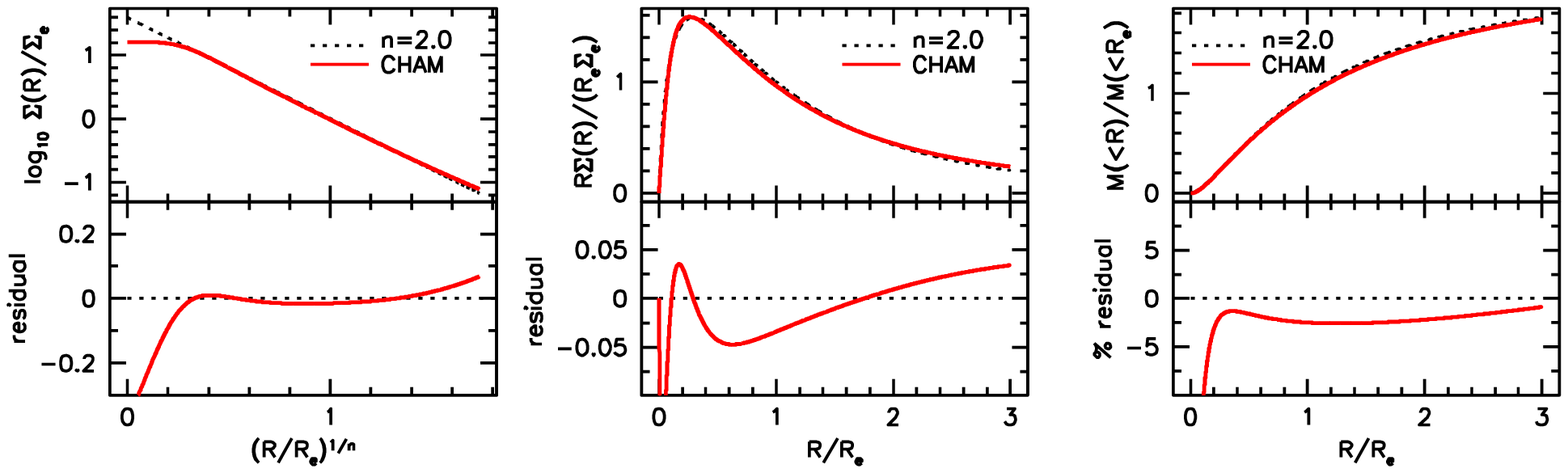}
\centering\includegraphics[width=0.99\linewidth]{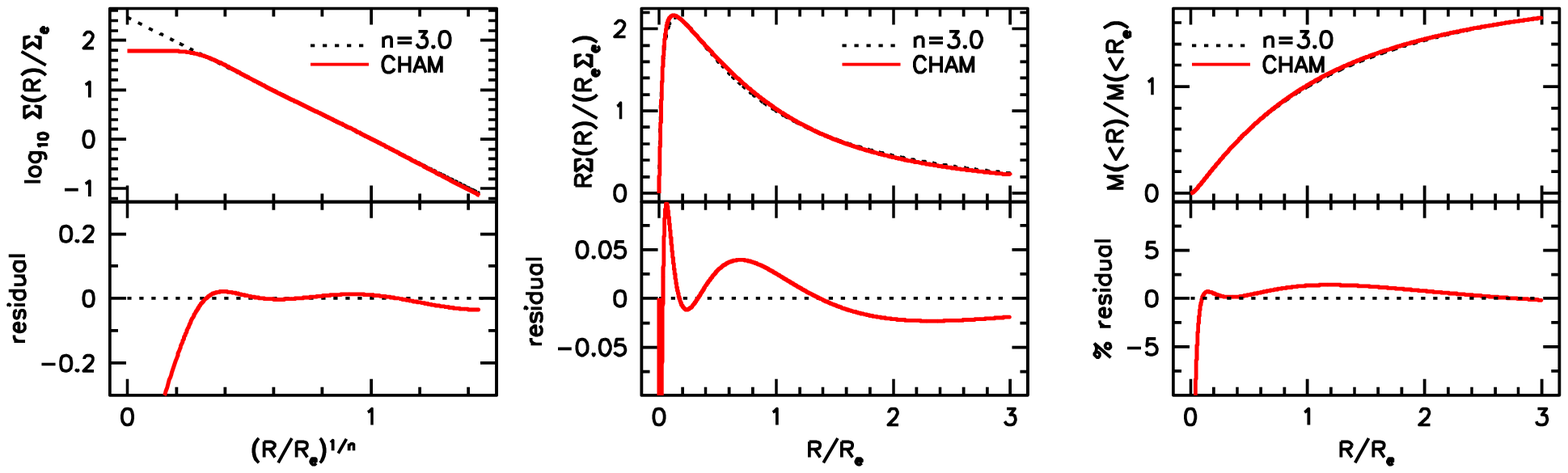}
\centering\includegraphics[width=0.99\linewidth]{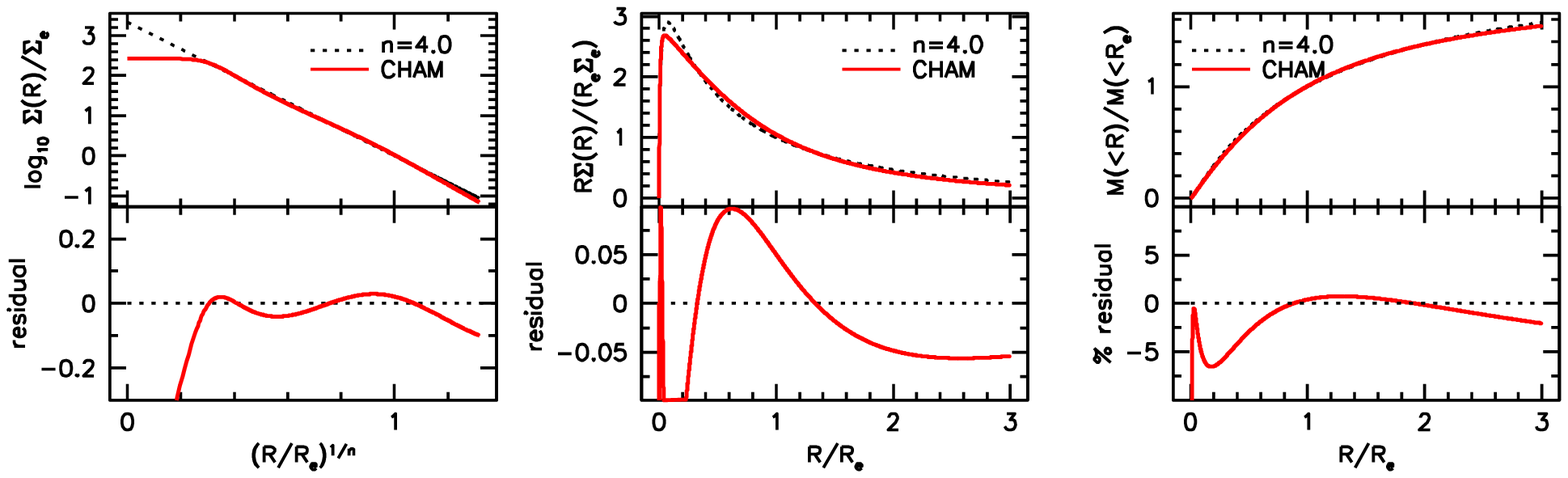}
\caption{Chameleon fits (red solid lines) to S\'ersic profiles (dotted
  lines) with $n=1,2,3,4$. The Chameleon profile reproduces the
  cumulative mass profile of a S\'ersic profile to a few percent over
  most radii interesting for strong lensing and kinematics.}
\label{fig:chmserfit}
\end{figure*}

\Fref{fig:chmserfit} shows Chameleon fits to S\'ersic profiles
with S\'ersic index $n=1,2,3,$ \& $4$ using the fitting function in
\Tref{tab:chmserfit}. The left panels show $\log_{10}\Sigma(R)$
vs $(R/R_{\rm e})^{1/n}$. In these units S\'ersic profiles are
straight lines. The Chameleon profiles deviate most significantly from
S\'ersic profiles at small radii, where the Chameleon profile has a
constant density core. The middle panels show $R\Sigma(R)$ vs
$R/R_{\rm e}$, which are the curves that were fitted against. The
right panels show the cumulative mass profile. For radii between $0.5
\lta R \lta 3.0$ effective radii the mass residuals are only a few
percent, and thus for most strong lensing and dynamical purposes the
Chameleon approximation to a S\'ersic profile is of sufficient
accuracy.

\begin{table}
 \centering
 \caption{Parameters of cubic fitting formula (\Eref{eq:power2})
   to the relations in \Fref{fig:chmser}.}
  \begin{tabular}{ccccccc}
\hline
\hline  
$x$ & $y$ & $x_0$ & $y_0$ & $y_1$ & $y_2$ & $y_3$ \\ 

\hline
$n$ &  $\log_{10}\left(\frac{\Sigma_{0,\rm Chm}}{\Sigma_{0,\rm Ser}}\right)$ & 1.69 & -0.254 & -0.259 & -0.036 & 0.014 \\
$n$ &  $\log_{10}\left(\frac{R_{0,\rm Chm}}{R_{0,\rm Ser}}\right)$ & 1.15 & 0.078 & -0.184 & 0.473 & -0.079 \\ 
$n$ &  $\log_{10}(\alpha_{\rm Chm})$ & 2.03 & -0.739 & -0.527 & -0.012 & -0.008 \\ 
\hline
\hline
\label{tab:chmserfit}
\end{tabular}
\end{table}

\label{lastpage}
\bsp

\end{document}